\documentclass[journal]{vgtc}

\onlineid{1803}

\vgtccategory{Research}

\vgtcpapertype{algorithm/technique}

\usepackage{tabu}      
\usepackage{booktabs}

\graphicspath{{./figs/}} 

\usepackage{amsfonts}
\usepackage{amsmath}
\usepackage{multirow}
\newcommand{\R}{\mathbb{R}}

\newcommand{\para}[1]        {\vspace{1pt}\noindent{\textbf{#1}}}

\newcommand{\etal}{{et al.}}

\usepackage{color}
\usepackage[dvipsnames,svgnames]{xcolor}
\usepackage{graphicx}
\usepackage{cite}
\usepackage{xspace}
\usepackage{verbatim}
\usepackage{amsmath}
\usepackage{amssymb}
\usepackage{url}
\usepackage{amsthm}
\usepackage[ruled,vlined,linesnumbered]{algorithm2e}
\usepackage{mathtools}
\usepackage{latexsym}
\usepackage{stmaryrd}
\usepackage{enumitem}

\usepackage{tabularray}
\usepackage{subfigure}
\usepackage{multirow}

\usepackage{makecell}

\newcommand{\Rspace}{\mathbb{R}}
\newcommand{\Xspace}{\mathbb{X}}

\newcommand{\Qcal}{\mathcal{Q}}

\title{Fast Comparative Analysis of Merge Trees \\ Using Locality Sensitive Hashing}

\author{
\authororcid{Weiran Lyu}{0009-0006-2887-174X}, 
\authororcid{Raghavendra Sridharamurthy}{0000-0001-8463-0488}, 
\authororcid{Jeff M. Phillips}{0000-0003-1169-2965}, and 
\authororcid{Bei Wang}{0000-0002-9240-0700}}

\authorfooter{
\item Weiran Lyu, Raghavendra Sridharamurthy, Jeff Phillips, and Bei Wang are with the University of Utah.
  	E-mail: wlyu@sci.utah.edu, g.s.raghavendra@gmail.com, jeffp@cs.utah.edu, beiwang@sci.utah.edu. 
}
\abstract{
Scalar field comparison is a fundamental task in scientific visualization. 
In topological data analysis, we compare topological descriptors of scalar fields---such as persistence diagrams and merge trees---because they provide succinct and robust abstract representations. Several similarity measures for topological descriptors seem to be both asymptotically and practically efficient with polynomial time algorithms, but they do not scale well when handling large-scale, time-varying scientific data and ensembles. In this paper, we propose a new framework to facilitate the comparative analysis of merge trees, inspired by tools from locality sensitive hashing (LSH). LSH hashes similar objects into the same hash buckets with high probability. We propose two new similarity measures for merge trees that can be computed via LSH, using new extensions to Recursive MinHash and subpath signature, respectively. Our similarity measures are extremely efficient to compute and closely resemble the results of existing measures such as merge tree edit distance or geometric interleaving distance. Our experiments demonstrate the utility of our LSH framework in applications such as shape matching, clustering, key event detection, and ensemble summarization.

}
\begin{document}
\maketitle
\section{Introduction}
\label{sec:introduction}

Measuring the similarity between objects is fundamental in data analysis.  
Particularly, it is important to quantify the proximity of objects to one another when they are notably similar, whereas measuring the distance between significantly dissimilar objects is often of lesser concern. 
A measure of similarity is the key to identifying repeated patterns, retrieving similar objects, building data clustering, and performing nearest neighbor search. 
It can also be used as a generalized inner product for kernel methods for a variety of embedding, classification, and  regression tasks.  
Choosing a meaningful similarity measure that can be computed efficiently is critical for performing advanced data analysis on any dataset, especially for visual analytics that  require real-time and interactive feedback. 

The need for meaningful and efficient similarity measures is especially true for scientific data analysis and visualization. 
In scientific computing, a large number of numerical simulations yield data in the forms of scalar fields, for example, temperature and surface atmospheric pressure from the Weather Research and Forecasting model. 
Moreover, the predominate way to facilitate efficient storage, analysis, and visualization of scalar fields is through various topological descriptors---from merge trees~\cite{CarrSnoeyinkAxen2003} to Morse--Smale complexes~\cite{EdelsbrunnerHarerZomorodian2003}---that represent the salient features of the underlying scientific phenomena. 

To that end, a rich set of comparative measures is available for a number of topological descriptors, with applications in structural biology, climate science, combustion studies, neuroscience, computational physics and chemistry, and ecology. 
A key takeaway from a survey by Yan \etal~\cite{YanMasoodSridharamurthy2021} is that choosing the right  similarity measure between topological descriptors is a recurring challenge because it tends to be both data-driven and dependent on specific applications. Rarely does a single measure fulfill all desired criteria, including acting as a metric or pseudometric, exhibiting stability and discriminative power, and being easy and efficient to compute.

The issue of computational efficiency is especially important in studying time-varying scientific data and ensembles at scale.   
For time-varying scalar fields, similarity measures between successive time steps are employed to identify periodic patterns, significant events, and anomalies, as well as to facilitate feature tracking (e.g.~\cite{SridharamurthyMasoodKamakshidasan2020, NarayananThomasNatarajan2015,SaikiaWeinkauf2017,SridharamurthyNatarajan2023}). 
Computational efficiency is particularly notable, for instance, for tracking the evolution of extreme weather events (e.g., thunderstorms and hurricanes) using time-varying reanalyzed data such as temperature, wind, and moisture (e.g.~\cite{WidanagamaachchiJacquesWang2017}). 
For ensembles, similarity measures aid in the identification of clusters, outliers, and distinctive ensemble members (e.g.,~\cite{GuntherSalmonTierny2014,SolerPlainchaultConche2018,YanWangMunch2020}). 
Efficient similarity measure holds particular significance in the examination of climate simulation ensembles, where thousands of climate model simulations with slight variations in parameter settings are utilized for climate projections. 
However, many similarity measures for topological descriptors suffer from challenges in both efficiency and scalability. While the complexity to compute the distance between a pair of merge trees of size $n$ is at best $O(n^2)$~\cite{YanMasoodSridharamurthy2021}, in practice, these methods involve solving matching problems, resulting in high runtime. Furthermore, in the case of ensembles or time-varying data, any useful analysis often requires a large number of comparisons, potentially involving all-pairs comparison in the worst case.   

The broader data analysis community has turned to locality sensitive hashing, or LSH~\cite{Charikar2002, IndykMotwani1998}, to address these questions of efficiency and scalability. LSH is more flexible than embedding methods which require intermediate vector representations, and it is less specific than clustering methods which typically force fixed groupings. 
LSH uses random hash functions, not to index distinct objects, but to randomly group together similar objects.  
Each object is given a set of representative signatures through a random process. 
These random signatures match between objects proportionally to how similar the objects are.  
This property induces a mechanism, whereby objects can be allocated to (multiple) hash buckets according to their signatures, enabling highly efficient probabilistic retrieval of similar objects by exploiting the tendency for them to reside within the same hash bucket. Importantly, LSH avoids comparing all objects.

Although LSH has become a fundamental tool in most large-scale data analysis, it remains unexplored in topological data analysis and visualization. 
As datasets continue to grow, it is paramount to investigate how to extend LSH to comparing topological descriptors at scale. 
This paper focuses on LSH for merge trees. 
The merge tree captures topological relations between sublevel sets of a scalar field, and is shown to be quite useful in symmetry detection~\cite{SaikiaSeidelWeinkauf2014}, shape matching and retrieval~\cite{SridharamurthyMasoodKamakshidasan2020}, feature tracking~\cite{SaikiaWeinkauf2017},   summarization~\cite{SridharamurthyMasoodKamakshidasan2020,YanWangMunch2020}, interactive exploration~\cite{PocoDoraiswamyTalbert2015}, and uncertainty visualization~\cite{YanWangMunch2020}.  
As most of the LSH algorithms require labeled structures, we  use labeled merge trees~\cite{YanWangMunch2020,YanMasoodRasheed2023}. A labeled merge tree is applicable when there is a natural labeling for the nodes, or when a labeling may be inferred from the data~\cite{LanParsaWang2023}. 
A climate simulation ensemble produces a set of slightly varying scale fields (e.g., pressure) that give rise to slightly different merge trees with a shared domain. 
We may use the indices of mesh nodes or the correspondences between underlying features (e.g., hurricane eyes) as the labeling, which is useful for feature tracking using merge trees (e.g.,~\cite{YanGuoPeterka2024}). A labeled merge tree where labels encode geometric features (e.g., Euclidean coordinates or other node attributes) further enables geometry-aware comparisons of merge trees~\cite{YanMasoodRasheed2023}.  

\para{Contributions.}
In this paper, we provide a LSH framework to facilitate the comparative analysis of labeled merge trees. 
Our framework is an adaptation of existing methods like Recursive MinHash~\cite{ChiLiZhu2014} and subpath signature~\cite{XuNiuJi2022} with critical extensions to our setting.  
The key contributions are as follows:
\begin{itemize}[noitemsep]
\item We propose two new similarity measures for labeled merge trees that can be computed via LSH, using new extensions to Recursive MinHash and subpath signature, respectively. 
\item Our similarity measures are extremely efficient to compute and closely resemble the results of existing measures such as merge tree edit distance or geometric interleaving distance.
\item  We develop efficient and scalable algorithms for our LSH framework in comparing labeled merge trees. 
\item Our experiments demonstrate the utility of our LSH framework in applications such as shape matching, temporal scalar field and ensemble summarization, and in identifying transitions between data structures of time-varying datasets.
\item We compare with existing methods in terms of accuracy and scalability. Our methods achieve $10-30\times$ speed-up on moderate datasets and high speed-up on a large ensemble.    
\end{itemize}
Overall, our framework is the first of its kind in integrating the notion of LSH within topological data analysis and visualization. It demonstrates good  efficiency in comparative analysis of merge trees at scale. Our LSH framework may be extended to handle other topological descriptors, in particular, extremum graphs~\cite{NarayananThomasNatarajan2015} and contour trees~\cite{CarrSnoeyinkAxen2003}.

\section{Related Work}
\label{sec:related-work}

\para{Merge trees.}
We mainly focus on merge trees and labeled merge trees; see the surveys~\cite{HeineLeitteHlawitschka2016,YanMasoodSridharamurthy2021} for other topological descriptors.

Merge trees capture the topology of sublevel sets of a real-valued function. 
They appear as an intermediate step in constructing contour trees~\cite{CarrSnoeyinkAxen2003}. In the past few decades, contour trees and merge trees have been used in various applications such as excess topology removal from isosurfaces~\cite{WoodHoppeDesbrun2004}, image analysis~\cite{MizutaMatsuda2005}, topology controlled volume rendering~\cite{WeberDillardCarr2007}, flexible isosurface generation~\cite{CarrSnoeyinkVanDePanne2010}, seed selection for segmentation~\cite{JohanssonMusethCarr2007}, high-dimensional data analysis~\cite{OesterlingHeineJanicke2011},  uncertainty data exploration~\cite{WuZhang2013}, cavity identification in biomolecules~\cite{BajajGilletteGoswami2009}, symmetry detection~\cite{ThomasNatarajan2014}, segmentation of volumetric data~\cite{BiedertGarth2015}, and analysis of astronomical data~\cite{RosenSethMills2021}. Multiple methods exist to compute contour trees/merge trees in both serial and parallel;  see~\cite{CarrSnoeyinkAxen2003,AcharyaNatarajan2015,CarrSewellLo2016,CarrWeberSewell2019,GueunetFortinJomier2016,GueunetFortinJomier2017,GueunetFortinJomier2019,WernerGarth2021}. 

Labeled merge trees were first defined by Gasparovic \etal~\cite{GasparovicMunchOudot2019}, followed by Yan \etal~\cite{YanWangMunch2020} and subsequently used by Yan \etal~\cite{YanMasoodRasheed2023} in comparing time-varying scalar fields. The nodes of a merge tree are labeled based on their function values or geometrical properties. 

\para{Comparative analysis of topological descriptors.} 
Comparison measures for topological descriptors such as merge trees and contour trees have to incorporate the structure along with the information about the scalar fields. 
Morozov \etal~\cite{MorozovBeketayevWeber2013} introduced interleaving distance between merge trees, a stable and discriminative distance but without an efficient algorithm to compute it. 
Later Beketayev \etal~\cite{BeketayevYeliussizovMorozov2014} introduced branch decomposition distance, which considers all possible branch decompositions. 
Sridharamurthy \etal~\cite{SridharamurthyMasoodKamakshidasan2020} introduced global merge tree edit distance for ordered and unordered trees extending tree edit distances, and provided an efficient algorithm to compute it. 
Sridharamurthy and Natarajan further extended it to local merge tree edit distance~\cite{SridharamurthyNatarajan2023} that enables a hierarchical comparison of merge trees. 

Gasparovic \etal~\cite{GasparovicMunchOudot2019} and Yan \etal~\cite{YanWangMunch2020} defined intrinsic interleaving distance for labeled merge trees, provided an algorithm to compute it with numerous applications. Pont \etal~\cite{PontVidalDelon2022} introduced Wasserstein distance between merge trees with the facility to compute barycenters and provided applications to ensemble data. 
Yan \etal~\cite{YanMasoodRasheed2023} introduced geometry-aware interleaving distance extending the intrinsic interleaving distance by incorporating geometric information to enhance its applicability to scientific data. Wetzel \etal~\cite{WetzelsLeitteGarth2022} followed by Wetzel and Garth~\cite{WetzelsGarth2022} introduced branch-decomposition independent edit distance and a deformation-based edit distance, both for comparing branch-decomposition of merge trees. Bollen \etal~\cite{BollenTennakoonLevine2023} introduced a stable edit distance addressing instability issues in previous edit distance-based measures. Wetzel \etal~\cite{WetzelsAndersGarth2023} eliminated the horizontal instability of edit distance-based measures by providing linear programming (LP) formulation of unconstrained edit distance. 
Qin \etal~\cite{QinFasyWenk2021} hashed persistence diagrams into binary codes using a generative adversarial network to speed up comparisons. While hashing is also used in our framework, we do not need any learning-based approach to generate representations of the merge trees. 
There is another set of comparison measures (such as those based on histograms~\cite{SaikiaSeidelWeinkauf2015,SaikiaWeinkauf2017} and the extended branch decomposition~\cite{SaikiaSeidelWeinkauf2014}). 
They are not metrics by definition but are simple, intuitive, and easy to compute. 

\para{Locality sensitive hashing.} 
Hashing is the process of transforming data to values, oftentimes of fixed size, via a hash function. 
It has been effective in compressing data for fast access and comparison; see~\cite{ChiZhu2017} for a survey. 
Locality sensitive hashing (LSH) utilizes a family of hash functions that map similar objects to the same hash buckets with high probability, making it useful for clustering and nearest neighbor search. 
Given a set of objects, LSH creates a sequence of discrete representatives for each object; the more these representatives match, the more similar the objects are deemed to be. 
Since the algorithm generates more representatives, it refines the notion of similarity, at the expense of higher computation cost. 
The basics of LSH were first introduced by Indyk and Motwani~\cite{IndykMotwani1998}, followed by Gionis \etal~\cite{GionisIndykMotwani1999}. Charikar~\cite{Charikar2002} and later Chierichetti \etal~\cite{ChierichettiKumarPanconesi2019} provided a more theoretical foundation along with necessary conditions for the existence of a LSH method for any similarity measure.  
 
Wu \etal~\cite{WuLiChen2020} provided a survey of LSH methods. These methods focus on evaluating a base measure---most commonly Jaccard similarity (MinHash), cosine similarity (SimHash), or string edit distance (SED). 
Ertl~\cite{ertl2020} provided an LSH framework for probabilistic Jaccard Similarity called ProbMinHash. Many LSH frameworks for SED have been defined, including by Zhang and Zhang~\cite{ZhangZhang2017,ZhangZhang2019,ZhangZhang2020}, Mar{\c{c}}ais \etal~\cite{MarccaisDeBlasioPandey2019}, Chen and Shao \cite{ChenShao2023}, approximating similarity search under SED by Mccauley~\cite{Mccauley2021}. 

Coming to structured data such as sequences, trees, and graphs, Wu and Li~\cite{WuLi2022} provided a survey of existing LSH methods. We discuss two relevant ones: hierarchical and kernel-based methods. 
Gollapudi and Panigrahy~\cite{GollapudiPanigrahy2008} proposed the idea of combining two min-hashes to enhance LSH for hierarchical structures like trees. 
Chi \etal~\cite{ChiLiZhu2014} provided Recursive MinHash (RMH) for hierarchical structures by repeatedly using MinHash at each level of the hierarchy, building the hash representations from the bottom up.

Kernel-based methods generate vectorized  representations of graphs while still being able to capture their structure. The most prominent kernel method for representing graphs and trees is the Weisfeiler-Lehman (WL) kernel by Shervashidze \etal~\cite{ShervashidzeSchweitzerVanLeeuwen2011}. Li \etal~\cite{LiZhuChi2012} defined Nested Subtree Hashing (NSH) which is an improvement over WL kernels, but still takes $O(n^2)$ time. 
Wu \etal~\cite{WuLiChen2017} defined $K$-ary Tree Hashing (KATH). All three methods, WL, NSH, and KATH kernels, although applicable to trees, are more suited for general graph structures with cycles. 
Aiolli \etal~\cite{AiolliDaSanMartinoSperduti2007} defined the Subset Tree (SST) kernel, which also takes $O(n^2)$ time. 
Tatikonda and Parthasarathy~\cite{TatikondaParthasarathy2010} defined a kernel for trees based on pivots and least common ancestors, but the computation of the kernel takes $O(n^2)$ time. 
Shin and Ishikawa~\cite{ShinIshikawa2018} and Xu \etal~\cite{XuNiuJi2022} defined subpath signature for trees to capture hierarchical relationships, providing an $O(n)$ signature construction. 
Signatures which asymptotically take $O(n^2)$ time are still useful for graphs, since graph comparison is a difficult problem to solve; however trees require signatures that are subquadratic to be of any use in practice.
Hashing techniques have been used for other applications such as graph alignment~\cite{HeimannLeePan2018} and relational graph matching~\cite{li2022}.

In terms of LSH methods, the upper bound and lower bound proofs for tree edit distance (TED)~\cite{GarofalakisKumar2003,GarofalakisKumar2005} along with the LSHability results~\cite{ChierichettiKumarPanconesi2019} showed that we cannot have an LSH method for TED.
Thus, we design an LSH method that emulates Jaccard similarity with some hierarchical information encoded so that the tree structure can still be incorporated. 
We concentrate on extending hierarchical methods by Chi \etal~\cite{ChiLiZhu2014} and some of the fast kernel methods such as subpath signature by Xu \etal~\cite{XuNiuJi2022} for their simplicity in implementation and potential for speed-ups.

\section{Background}
\label{sec:background}

\subsection{Merge Trees and Labeled Merge Trees}
\label{sec:merge-trees}
\para{Merge trees.} 
Given a scalar field $f: \Xspace \to \Rspace$ defined on a topological space $\Xspace$, the merge tree of the data $(\Xspace,f)$ captures the connectivity of its sublevel sets $f^{-1}(-\infty, t]$ (for some $t \in \Rspace$). 
Mathematically, we identify two points $x,y \in \Xspace$ to be equivalent (i.e.,~$x \sim y$) if $f(x) = f(y)=t$  and they belong to the same connected components of $f^{-1}(-\infty, t]$. 
The \emph{merge tree} is the quotient space $\Xspace/\sim$. 
The root of a merge tree is the global maximum, the leaves are the local minima, and the internal nodes are the merging saddles.
We denote a merge tree by $T$, with a set of nodes $V(T)$ and a set of edges $E(T)$.

\begin{figure}[!ht]
\centering
\vspace{-3mm}
\includegraphics[width=0.9\columnwidth]{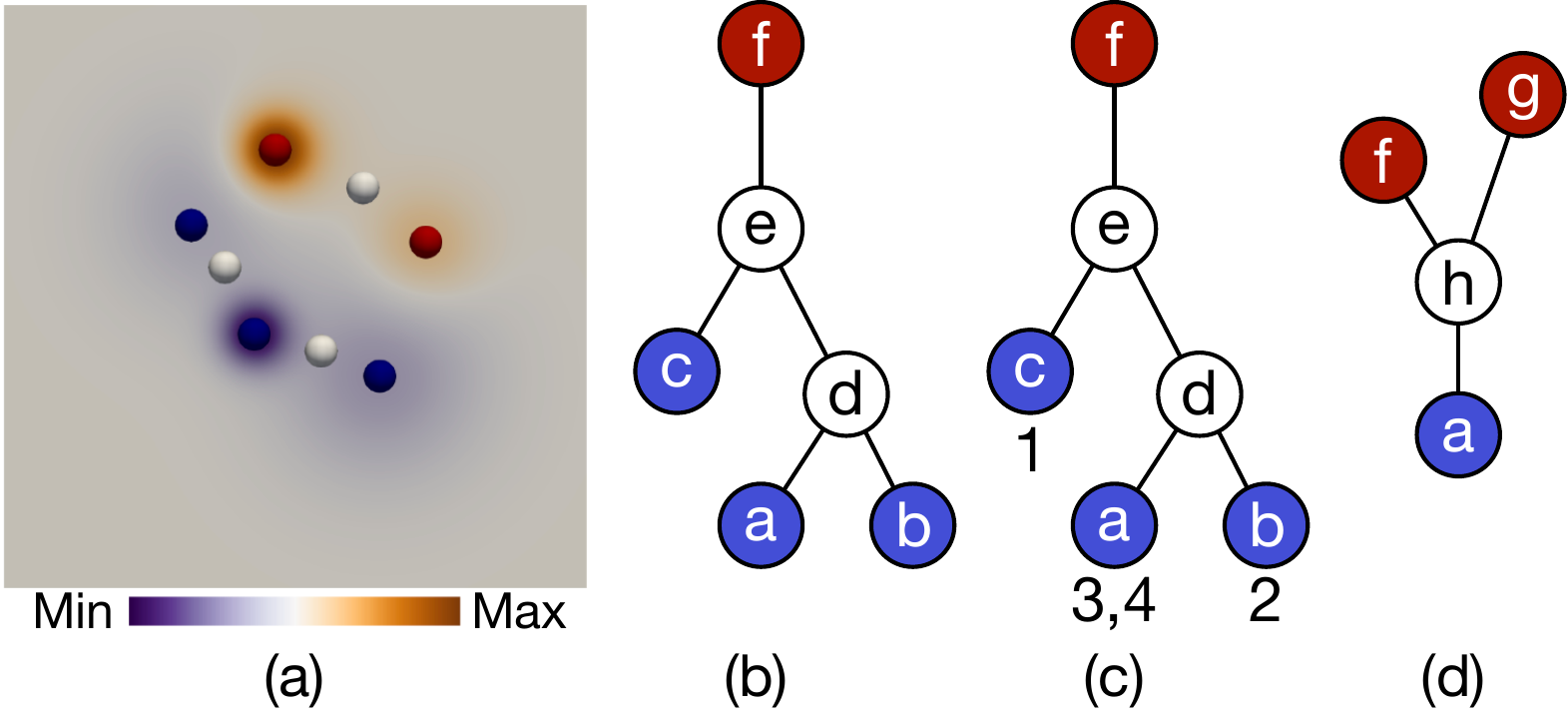}
\vspace{-3mm}
\caption{Illustration of merge trees. (a) A scalar field $f$ overlaid with critical points. (b) Merge tree. (c) Labeled merge tree with node labels  $[3,4]$, $[2]$, and $[1]$ for nodes a, b, and c, respectively. (d) Merge tree of $-f$.}
\label{fig:merge-tree}
\vspace{-4mm}
\end{figure}

\para{Labeled merge tree.}
A labeled merge tree consists of a merge tree $T$ with a map $\pi:[n] \to V(T)$ that is surjective on the set of leaves~\cite[Def. 2.2]{GasparovicMunchOudot2019}, that is, all leaves and some internal nodes are labeled. Here, $[n]:=\{1,2,\dots,n\}$ denotes a set of labels.   
\cref{fig:merge-tree} shows merge trees and labeled merge trees for a 2D scalar field.

A node can have multiple labels. 
Labels can be added from one labeled merge tree to another unlabeled merge tree using a variety of methods~\cite{YanMasoodRasheed2023}: 
\emph{tree mapping} considers topological information captured by how the function values differ along the path containing the least common ancestor; \emph{Euclidean mapping} considers geometric information captured by the Euclidean distance between the critical points; and \emph{hybrid mapping} combines the two.

\para{Merge tree edit distance ($d_E$)}. 
Given two merge trees $T_1$ and $T_2$, let $\Qcal:=\Qcal(T_1, T_2)$ denote a set that contains sequences of edit operations (insert, delete, relabel) that transform $T_1$ to $T_2$.  
$\gamma(Q)$ denotes the cost incurred over a sequence $Q \in \Qcal$; it is the sum of the cost of individual edit operations.  Now, $d_E$ is given by
$
d_{E}(T_1,T_2) = \min_{Q \in \Qcal} (\gamma(Q)).
$

\para{Geometry-aware interleaving distance ($d_I$)}. 
For $T$ with a labeling $\pi$ that incorporates both geometric and topological information, and a scalar function $f$, its \emph{induced matrix} $M$ consists of entries defined as $M_{ij}=f(lca(\pi(i),\pi(j)))$ where $lca$ stands for the \emph{lowest common ancestor}. Given two labeled merge trees $T_1$ and $T_2$, $d_I$ is given by the \emph{cophonetic metric} (i.e.,~$p$-th norm)~\cite{CardonaMirRossello2013} between the induced matrices $M_1$ and $M_2$, i.e.,  $d_I(T_1,T_2) = \lVert M_1 - M_2 \rVert_p.$ 
We use $p=\infty$. 

\subsection{Locality Sensitive Hashing}
\label{sec:LSH}
We define LSH and its variants that are relevant to this work. For a more detailed introduction, see the survey by Wu~\etal~\cite{WuLiChen2020}.

\para{Locality sensitive hashing.} 
An LSH algorithm considers a space of objects $\mathcal{Z}$ and a similarity function $s : \mathcal{Z} \times \mathcal{Z} \to [0,1]$. 
It then uses a family of hash function $\mathcal{H}$, so that in expectation
\[
\mathbb{E}_{h \sim \mathcal{H}} [h(p) = h(q)] = s(p,q). 
\]
In other words, given the randomness in the choice of harsh functions $h \sim \mathcal{H}$, the probability that $h(p) = h(q)$ equals their similarity $s(p,q)$~\cite{IndykMotwani1998}. Therefore, the greater the similarity between objects $p$ and $q$, the higher the probability that they will hash together. This is in contrast to traditional hashing used for indexing a set, where we aim to avoid collisions between objects. 

Then we bundle a set of hashes to amplify their effect, so objects more similar than a threshold collide with high probability. We harness this property to compute similarity or find similar objects without explicit comparison or search.

\begin{figure}[!ht]
\centering
\vspace{-2mm}
\includegraphics[width=1.0\columnwidth]{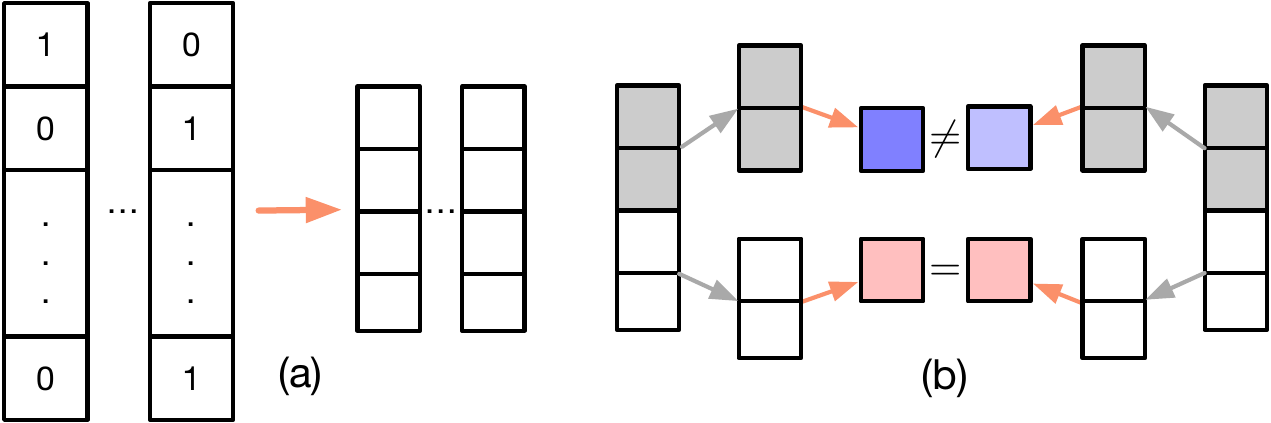}
\vspace{-5mm}
\caption{(a) Generating MinHash: an $N \times M$ binary matrix is generated and then reduced to a dense $q \times M$ signature matrix using $q$-MinHash. Here, $q=4$. (b) LSH:  a signature is divided into bands of length $r$. Here,  $r=2$. LSH is applied to individual bands and candidate pairs are determined based on signature collision in any band (in red). The orange arrows are applications of hash functions.}
\label{fig:LSH}
\vspace{-4mm}
\end{figure}

\para{MinHash.} 
A MinHash is a common mechanism to create an LSH when each object $Z \in \mathcal{Z}$ is represented as a subset $Z \Rightarrow W \subset U$; where $U = \{u_1, \ldots, u_N\}$ is some universal set of possible objects.  
The base element of a MinHash is a (random) permutation over $\sigma : U \mapsto U$.  However, the actual hash $h$ returns only the minimum $h(W) = \arg \min_{w \in W} \sigma(w)$. As a result, we typically do not implement this as a full permutation (which can be expensive to store and compute), but $\sigma$ can map $U \to \R$, so the minimum object  can still be retrieved.  

A classic insight~\cite{BroderCharikarFrieze1998} is that for a MinHash $h$ on two sets $V, W \subset U$, 
\[
\mathbb{E}_{h \sim \mathcal{H}} [h(V) = h(W)] = \mathsf{Jac}(V,W),
\]
where $\mathsf{Jac}(V,W) = {|V \cap W|}/{|V \cup W|}$ is the Jaccard similarity.  

Therefore, to use a MinHash for any family of objects $\mathcal{Z}$, the key is to represent each $Z \in \mathcal{Z}$ into sets $W$ of potential ``views'' or structured ``subset-elements'' of those base objects $Z$.  Then, we can directly invoke MinHash to define a similarity of those base objects via the Jaccard similarity of their sub-objects.  

We combine $q$ MinHash functions $h_1, \ldots, h_q$ into a single hash function called an $q$-MinHash.  For a set $W$, it concatenates these $q$ signatures into a longer ordered set $[h_1(W), h_2(W), \ldots, h_q(W)]$; see \cref{fig:LSH} for an illustration.

\para{LSHability}~\cite{ChierichettiKumarPanconesi2019,Charikar2002} provides criteria for the existence of an LSH framework for a particular similarity measure. For a similarity $s$ to admit an LSH framework, there are two necessary conditions:
\begin{enumerate}[noitemsep,leftmargin=*]
    \item $1-s$ must be a metric distance function;
    \item $1-s$ must be isometrically embeddable in $l_1$.
\end{enumerate}
For an LSH algorithm to exist, both the distortion lower bound and upper bound should be $1$ according to~\cite{ChierichettiKumarPanconesi2019}. 
Garofalakis and Kumar proved that this is not the case for tree edit distance~\cite{GarofalakisKumar2003,GarofalakisKumar2005}. 
Thus, in this work, we operate with the Jaccard similarity variants.

\para{Amplifying similarity for similarity search.} 
A single hash that has collisions proportional to object similarity by itself would make for a very noisy similarity search. For objects with a very high similarity (e.g.,~$0.9$), they do not collide $10\%$ of the time. To address this issue, we use a set of $k$ randomly chosen hash functions $h_1, \ldots, h_k \sim \mathcal{H}$ and employ \emph{banding}.

We divide the $k$ hash functions into $b$ \emph{bands} each with $r = {k}/{b}$ hash functions.  
A band is considered to have a collision for two objects  only if \emph{all} $r$ hash functions in the band show a collision. In other words, these $r$-banded hash functions are combined into a single $r$-MinHash function. 
Such a process can be implemented efficiently by concatenating the hash values from each $h_i$ in the band, and using a regular hash table to increase the specificity, thereby making it harder for objects to collide. 

However, we counterbalance this increased specificity by decreasing it through $b$ bands, where two objects are deemed to collide if they collide in \emph{any} band. Together, these adjustments sharpen the threshold for identifying a collision. Consequently, pairs with $s(p,q)=0.9$ are much more likely to be identified as being similar, and pairs with $s(p,q)=0.1$ are much less likely to be considered similar.  

For a fixed $k$, we can adjust the desired specificity and the similarity threshold by changing $r$.  In our case, we find that the hash function design is already very specific, so we use a very small $r$.  
\cref{fig:LSH} provides an illustration of conceptual steps of how LSH is used to identify similar objects  efficiently. Objects $W_i$ and $W_j$ are considered similar when their hashed signatures collide in any band. 

\subsection{Subpath Signature for LSH}
For graph-structured data, a common way to encode labeled graphs into sets $W$ from some universal set $U$ is to consider all subpaths. Given a parameter $t$, we set $W$ as all sequences of $t$ labels which can be formed by a path in the graph. The representation of an object as a set $W$ means it can immediately be used in the MinHash framework.   
For rooted trees, Xu \etal~\cite{XuNiuJi2022} showed that it is effective to consider only directed paths from the root toward the leaves, without including the root itself.   

A rooted-tree subpath signature is of interest because it not only is efficient but also provides theoretical bounds w.r.t. to the tree edit distance $d_E$ with unit costs (see~\cite[Theorem 2]{XuNiuJi2022}). 
Specifically considering $d_E$ with operations consisting of insertion, deletion, relabelling, and subtree moves, if the subpath signatures of two trees for parameter $t$ are the same, i.e., ~$S_t(T_1) = S_t(T_2)$, then 
\[
d_E(T_1,T_2) \le n - \min (t-1, \mathrm{height}(T_1)+1), 
\]
where $n:=|T_1|$ $(n>2)$ and $t > 1$.

\subsection{LSH for Hierarchical Data}
\label{sec:RMH}
In this section, we describe how LSH is used to specifically handle hierarchical data such as documents and fixed-height trees.

\para{Recursive MinHash (RMH)}~\cite{ChiLiZhu2014} uses $q$-MinHash repeatedly by following a bottom up approach to compare hierarchical data.
\begin{itemize}[noitemsep,leftmargin=*]
    \item The method \textbf{recursively} performs the following steps starting from the lowest level, until the highest level is reached:
    \begin{enumerate}
        \item Apply \textbf{$q$-MinHash} to data at the current level.
        \item \textbf{Reorganize} these $q$-MinHash vectors into $q$ sets: insert the $i$th term from each $q$-MinHash vector into the $i$th set. 
    \end{enumerate}
    \item Finally, we apply $q$-MinHash again to the reorganized vectors and \textbf{concatenate} the results into one single fingerprint.
\end{itemize} 

\begin{figure}[!ht]
\centering
\vspace{-4mm}
\includegraphics[width=0.8\columnwidth]{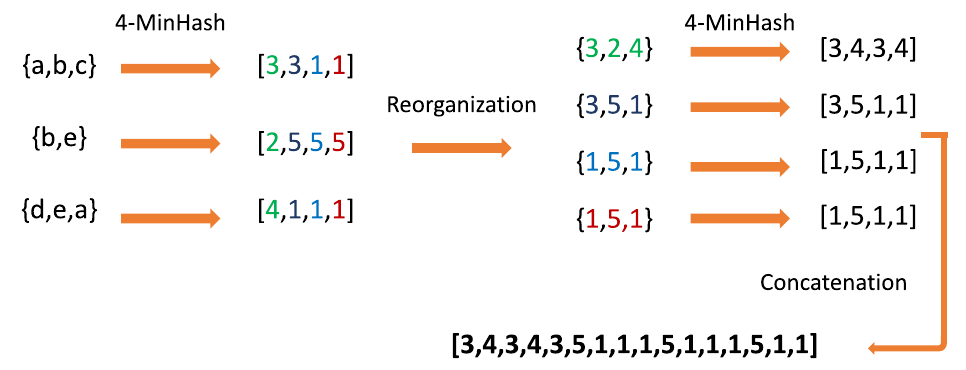}
\vspace{-4mm}
\caption{An illustration of RMH.}
\label{fig:text-rmh}
\vspace{-4mm}
\end{figure}

\cref{fig:text-rmh} shows how RMH works for a toy example consisting of two levels, with $q=4$. The sets $\{{a,b,c}\}, \{{b,e}\}, \{{d,e,a}\}$ comprise the bottom level. First, $q$-MinHash is applied on these sets, and the hashes are reorganized. Next, $q$-MinHash is generated for the reorganised sets to account for the top level. The final signature is the ordered set of these $q$-MinHash signatures for these reorganized sets. We describe the hash functions in detail in the supplement.

A single fingerprint is represented as an ordered set of $k$ values. These values can then be split into $b$ bands of $r=k/b$ elements, each within an LSH framework. The LSH parameters $r$ and $b$ do not need to correspond with the signature parameter $q$.  

\section{LSH for Comparing Merge Trees}
\label{sec:method}

Our LSH framework introduces two similarity measures: one based on Recursive MinHash $s_R$, and another based on subpath signature $s_S$. 
To compare against distance matrices in our experiments, we convert each similarity measure to a distance, denoted as $d_R$ and $d_S$, respectively, where $d_R = 1 - s_R$ and $d_S = 1 - s_S$. We present an overview of our framework followed by a detailed description of the algorithms. 

\begin{figure}[!ht]
\centering
\vspace{-4mm}
\includegraphics[width=1\columnwidth]{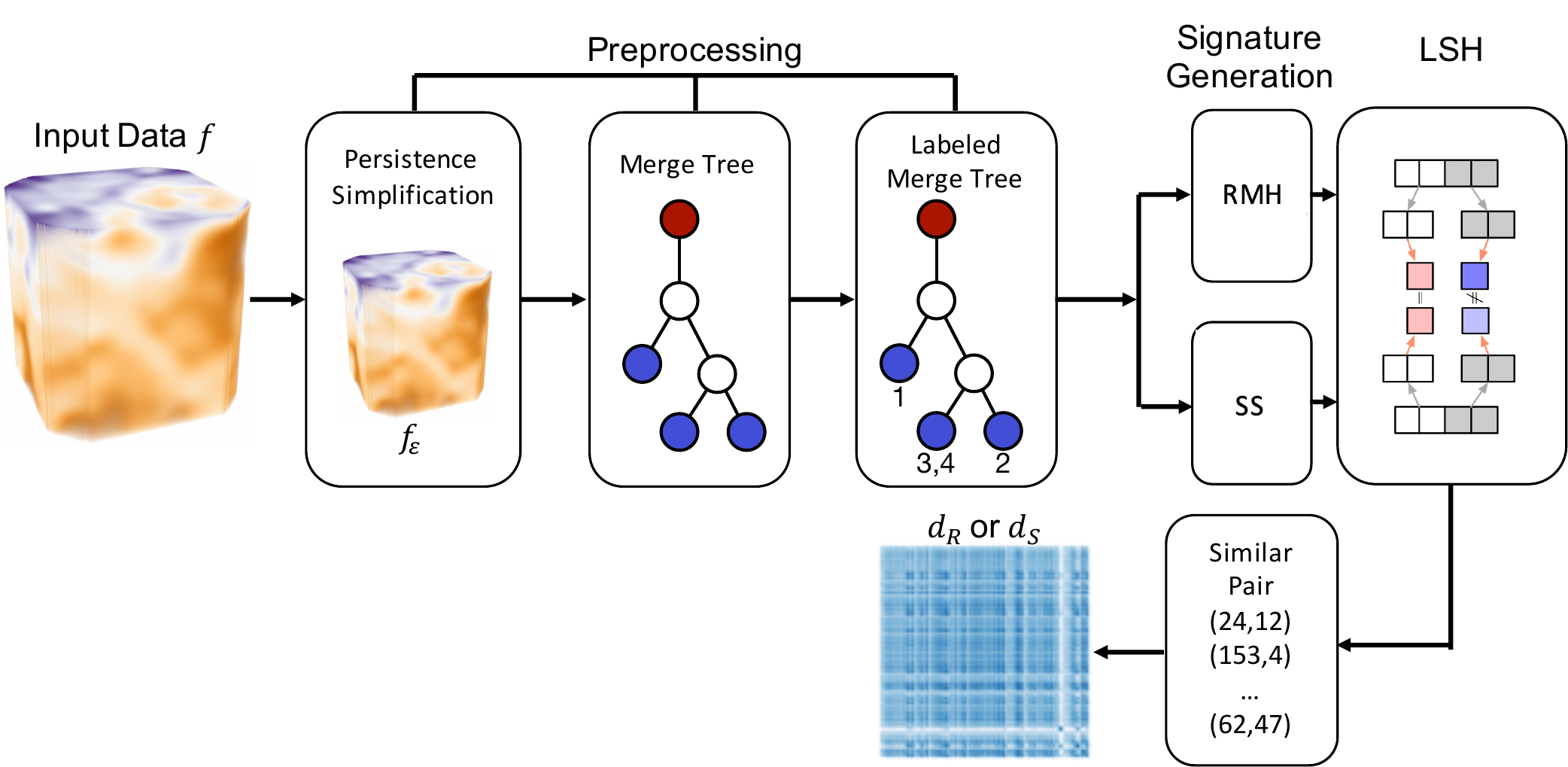}
\vspace{-4mm}
\caption{A pipeline of LSH for comparing merge trees.}
\label{fig:pipeline}
\vspace{-4mm}
\end{figure}

\para{Overview.} An overview of our pipeline is shown in \cref{fig:pipeline}. Given a set of scalar fields as input, we generate our $d_R$ and $d_S$ matrices as follows:
\begin{itemize}[noitemsep,leftmargin=*]
\item \para{Preprocessing.} We first simplify each scalar field with a small persistence threshold to remove noise in the data~\cite{EdelsbrunnerLetscherZomorodian2002}. We then compute its corresponding merge tree, followed by label assignment. 
\item \para{Signature generation.} We take the labeled merge trees as the input and generate their signatures using either RMH or subpath signature (SS) algorithms. 
\begin{itemize}[noitemsep,leftmargin=*]
\item SS signature: For each labeled merge tree, we can identify the elements $W$ of a universal set $U$ of subpaths. Then, we directly use  MinHash to obtain signatures.
\item RMH signature: We combine RMH with a hierarchy of trees, and apply it to labeled merge trees to generate signatures. 
\end{itemize}
\item \para{LSH.} We divide signatures into $b$ bands each with $r$ rows. If two objects collide in any band, then we mark these two objects as a similar pair. For empirical comparison, we generate our distance matrices $d_R$ and $d_S$ by collecting all similar pairs from the LSH.
\end{itemize}

\subsection{Subpath Signature for Labeled Merge Trees} 
We extend the rooted-tree subpath signature method to labeled merge trees.  
Given labeled merge trees and a parameter $t$ that denotes subpath length, we generate all subpaths of length $t$ and collect them in a multiset. 
We include a path containing $t-1$ dummy nodes leading to the root node. This ensures that all the subpaths would be of length $t$, even in the extreme case where the tree consists of only one node. 

\begin{figure}[!ht]
\centering
\vspace{-2mm}
\includegraphics[width=0.8\columnwidth]{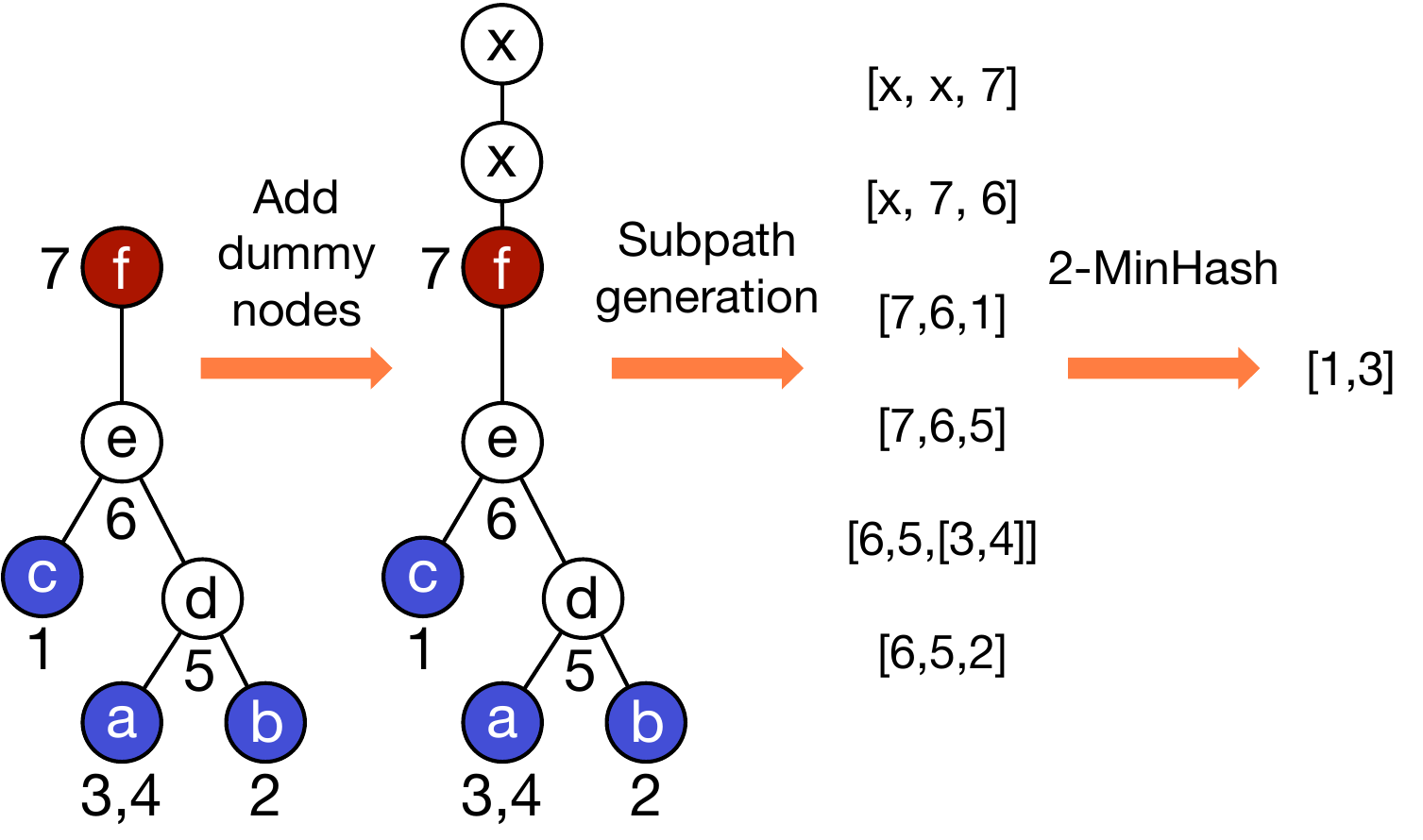}
\vspace{-2mm}
\caption{Illustration of subpath signature for a labeled MT.}
\label{fig:ss}
\vspace{-4mm}
\end{figure}

\cref{fig:ss} shows an example on generating subpath signatures for a labeled merge tree with $t=3$ and $q=2$. After adding two dummy nodes, we start collecting all the subpaths of length three from the new root and traverse down the tree until every node has been visited. We then apply $2$-MinHash to each merge tree to get a collection of vectors. Finally, we take these vectors as input for LSH.

We design a simple modification of depth-first search (DFS) to generate subpaths. Intuitively, a traditional DFS returns a subpath of length $1$.
To modify it to return a subpath of length $t$, we do the following steps: 
whenever a node is flagged as visited in a traditional DFS, it is popped and returned. 
Instead of a single node, we implement a $(t,t-1)$ pop-push operation that pops $t$ elements to provide us with the subpath and immediately pushes back $t-1$ elements that have not been flagged as being visited. 

In practice, if we implement the stack using an array, then we need not explicitly perform $(t,t-1)$ pop-push operations since we can access the elements that are not at the top of the stack. Thus, we generate the multiset containing all the subpaths of length $t$. 
The supplement provides the pseudocode.  
\cref{algorithm:signature} generates the subpath multiset. 
\cref{algorithm:signature-visit} is the modified DFS that is used internally by~\cref{algorithm:signature}. 
It modifies traditional DFS with the extra step $\mathsf{POP}-\mathsf{PUSH(t,t-1)}$, which takes $O(t)$ time.  
Thus, the running time is given by $O(n \times t)$ where $n$ is the size of the tree. 
Since $t \ll n$, and $t < 10$ is a constant in practice for most cases, the running time is linear in the size of the trees.  

\subsection{RMH for Labeled Merge Trees}
\label{sec:dR}

Our starting point is RMH, as described in~\cref{sec:RMH}.  RMH provides a hash signature for hierarchical structures.  It was originally designed for documents where the hierarchy is fixed, consisting of words, sentences, paragraphs, and so on. 
However, for merge trees, the hierarchy (the tree structure) can be different for each object as it depends on the topology of the sublevel sets. 

We first make the new observation that the RMH framework can be applied to any hierarchical object.  
Every reorganization step results in $q$ sets.  Then, in the next recursive step, $q$ sets always lead to $q$ different $q$-MinHash signatures.  This is true no matter how many iterations of the hierarchy have been processed. Hence, the final concatenated signature is always of length $k=q^2$.  Since the above process is independent of the height or structure of the tree, the signature can be compared across trees of different topology.  

Second, unlike the traditional RMH where all nodes at a particular level are of the same type, in the case of merge trees, the nodes can be of different combinations. 
At a particular level, the nodes can be all extrema, a mixture of extrema and saddles, all saddles, or the root node. To address the combination of different nodes, 
we apply a recursion based on the type of the nodes rather than the level. This ensures that the MinHash of subtree rooted at $a$ and the MinHash of subtree rooted at $b$ will be subjected to reorganization only at the level corresponding to their lowest common ancestor, $lca(a,b)$. 
We modify the RMH to reflect this small but crucial difference so that it can be applied to merge trees; see \cref{algorithm:dR} in the supplement for the pseudocode.  

In \cref{fig:rmh}, we provide a toy example illustrating how RMH with $2$-MinHash works for labeled merge trees. First, we apply $2$-MinHash to all the labels of leaf nodes in our merge tree. Second, we reorganize the hash vectors of leaf nodes that share the lowest common ancestor, leaf nodes $a$ and $b$, and their parent node $d$. Third, we truncate the tree to move one level up. We perform the three steps above  recursively until we only have one node left, which is the root. Then, we apply $2$-MinHash again to obtain the hash vectors and concatenate them into a single RMH signature as the input for LSH.

\begin{figure}[!ht]
\centering
\vspace{-3mm}
\includegraphics[width=1.0\columnwidth]{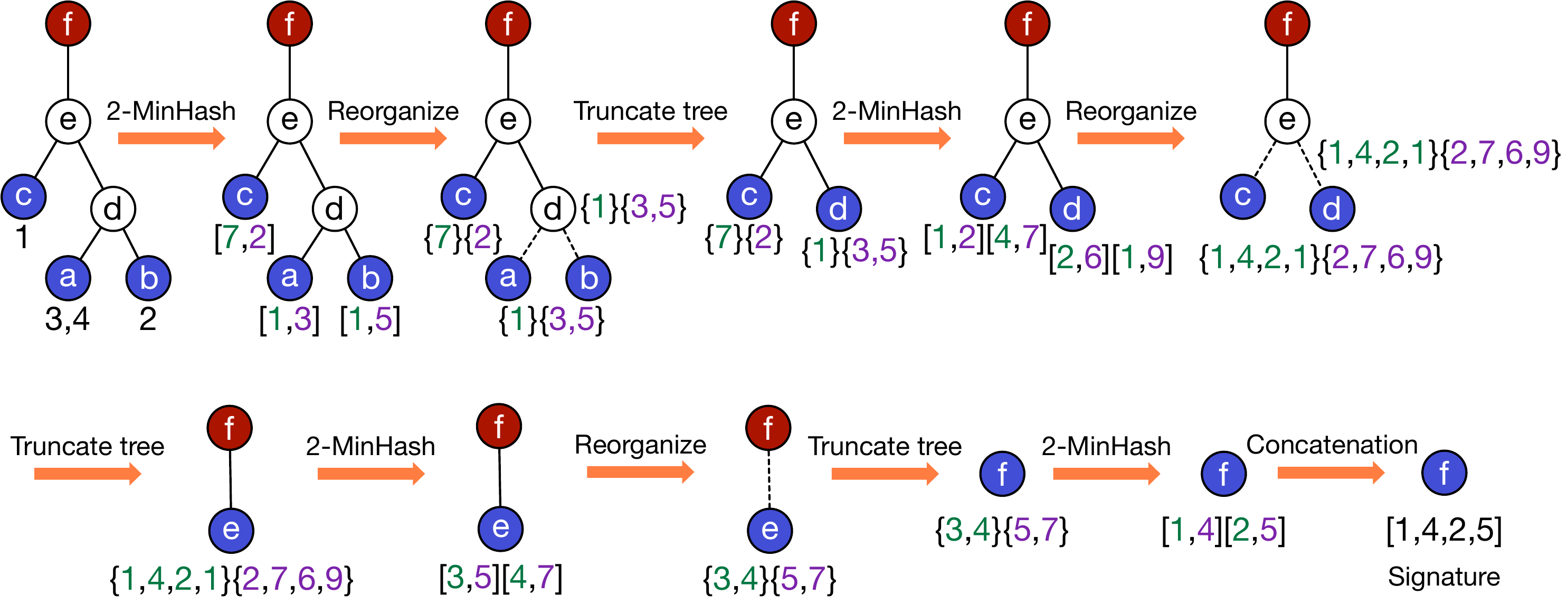}
\vspace{-4mm}
\caption{An illustration of RMH on a labeled MT.}
\label{fig:rmh}
\vspace{-4mm}
\end{figure}

Using our observation that each recursive round always generates at most $q$ sets, we provide a simplified analysis of the runtime compared to \cite{ChiLiZhu2014}.  
We say a node is an \emph{exposed leaf node} if it has not yet been truncated in the recursive process, and it has no child nodes in the truncated tree.  In each round, each exposed leaf node needs to convert from at most $q$ sets, each of size at most $q^2$, to $q$ MinHash signatures. This process takes $O(q^4)$ time.  For binary trees of height $L$ with $n$ nodes, there are at most $\min{(2^L,n)}$ exposed leaf nodes.  Since the height decreases by one each round, the total runtime is 
\[
\sum_{z=1}^L O(\min{(2^{L-z+1},n)} q^4)) = O(L n q^4).
\]
Therefore, if $q$ is a small constant (we typically use $q=2$ or $4$), then the runtime is near-linear in the size of the tree unless it is very unbalanced and the height $L$ is super-logarithmic in the tree size.  

\begin{table*}[]
    \centering
    \caption{Detailed descriptions of all datasets and parameters used in all experiments. $\varepsilon$ stands for persistence threshold.}
    \vspace{-2mm}
    \begin{tabular}{c|c|c|c|c|c|c} 
        \hline
        Dataset & Dimensions & \# Instances/Time steps & $\varepsilon$ & Size of Merge Tree & $k$ & $r$ \\
        \hline
        Vortex Street & $192 \times 64 \times 48$  & 102 & 0.01 & 10 - 104 nodes & {20, 40, 60, 80} & {1, 2} \\
        Shape Matching & varies & 132 & 0.01 & 10 - 52 nodes & {20, 40, 60, 80} & {1, 2, 4} \\
        Corner Flow & $450 \times 150$ & 1500 & 0.01 &  12 - 78 nodes & {20, 40, 60, 80} & {1, 2} \\
        Heated Flow & $150 \times 450$ & 2000 & 0.01 & 2 - 148 nodes & {20, 40, 60, 80} & {1, 2} \\
        Viscous Finger & $100 \times 100 \times 100$ & 5746 & 5 & 64 - 232 nodes & {20, 40, 60, 80}& {1, 2} \\
        \hline
    \end{tabular}
    \label{table:data}
    \vspace{-4mm}
\end{table*}

\subsection{Design Choices and Implementation Details}

Labeled merge trees may contain nodes with multiple labels. 
To ensure that we do not miss any similar pairs, we incorporate multiple labels into our subpath, i.e.,~if node $a$ has $l$ labels then we generate a subpath which contains $a$, $b$, and $c$ as $((a_1, \ldots, a_l), b, c)$. 

We follow the traditional LSH method that the signatures are divided into $b$ bands of $r$ rows each. A ``match'' between merge trees requires that for at least one band, all its rows have to match. Therefore, if $r$ is larger, we require a closer match, and if $b$ is larger, it is more forgiving as there are more chances to find a instance of all-rows correspondence. Since the relationship $k = r \times b$ is fixed, then for a specified $k$, we can tune the similarity threshold to define a ``close pair'' to be more specific by increasing $r$; so adjusting $r$ is like adjusting a threshold for similarity with any other distance. As we have observed, small $r$ values (often $r=1$) provide useful matches already.  

We can also adjust the value of $k$: the larger the parameter $k$, the less variance in the matches found in this randomized process, but the more expensive the computation. We discover via experiments that a fairly moderate value of $k$ (e.g., $k=20$) works quite well.  

If we have a very large number of trees, an additional hash function can be applied to report collisions within a single band. However, given the scale of our experiments (not involving millions of trees), it is feasible to calculate and report all collisions directly. This approach enhances the precision of our reporting for this paper.

\para{Implementation details.}
We use the implementation in~\cite{YanMasoodRasheed2023} to generate merge tree labels. 
We implement both $d_R$ and $d_S$ in Python. 
We also implement the algorithm to generate subpaths of length $t$ required for $d_S$. 
We implement the MinHash first introduced by Broder~\cite{Broder2000}.

\section{Experiments and Results}
\label{sec:result}

We experimentally validate the effectiveness of our LSH framework, in terms of \textbf{utility} and \textbf{efficiency}: 
\begin{itemize}[noitemsep,leftmargin=*]
\item We demonstrate that our new similarity measures are effective on a wide variety of examples, in recovering results from existing merge tree distances, and in some cases uncovering new scientific structures. 
\item We illustrate that our framework is significantly more efficient than standard distance measures.~It achieves $10$-$30\times$ speed-up on moderately-sized datasets, and (estimated) $800\times$ on large ones. 
\end{itemize}

\para{Comparison with edit and interleaving distances.}
Our framework introduces two similarity measures: one based on Recursive MinHash $s_R$ and another based on subpath signature $s_S$.
We convert each similarity measure to a distance, denoted as $d_R$ and $d_S$, respectively, where $d_R = 1 - s_R$ and $d_S = 1 - s_S$. 

We compare our results with the merge tree edit distance~\cite{SridharamurthyMasoodKamakshidasan2020} (referred to as edit distance for short, denoted as $d_E$) and  the geometry-based interleaving distance~\cite{YanMasoodRasheed2023} (denoted as $d_I$). 
We choose these two distances as they are both applicable to labeled merge trees but represent different types of similarity measures. 
$d_E$ provides the best theoretical runtime among comparative measures that internally solve matching problems, whereas $d_I$ decouples the   computation externally into a labeling step and a comparison step. 

In our comparison, we report $d_E$ and $d_I$, and use a color scale to visually observe the thresholds when useful structures become apparent. 
We then demonstrate that our LSH framework could approximately match these structures as follows: for each fixed $k$, we adjust $r$ so that the matched merge trees approximately correspond with those from the other distances.  
We visualize this optimal choice of $r$ as a symmetric binary matrix.  
In some cases, the other distances do not find an interesting structure, but our LSH-based measures do. 

We use the Topology Toolkit (TTK)~\cite{TiernyFavelierLevine2018} to visualize scalar fields and generate merge trees. 
We compute $d_E$ and $d_I$ using implementations provided in~\cite{SridharamurthyMasoodKamakshidasan2020} and~\cite{YanMasoodRasheed2023}, respectively. 
We use a hybrid mapping strategy from~\cite{YanMasoodRasheed2023} to generate labels for all datasets, since it encodes both geometric and topological information from the input. 
We perform our experiments on a standard laptop with {i7} processor with $20$ threads running at $3.5$ GHz, with $32$ GB memory. We use Moving Gaussian, a toy dataset from~\cite{YanMasoodRasheed2023}, TOSCA shape dataset from~\cite{TOSCA2024}, Corner flow~\cite{Baeza-RojoGunther2020} and Heated flow~\cite{GuntherGrossTheisel2017} datasets from~\cite{gerrisflowsolver} (available at~\cite{Cglflow2024}), 3D vortex street dataset from~\cite{CamarriSalvettiBuffoni2005} (available at~\cite{3Dvs2024}) and Viscous Fingers dataset~\cite{Viscousfingers2016}.

\subsection{An Illustrative Example}
\label{sec:Gaussian}

\begin{figure}[!ht]
\vspace{-3mm}  
\centering
\includegraphics[width=1.0\columnwidth]{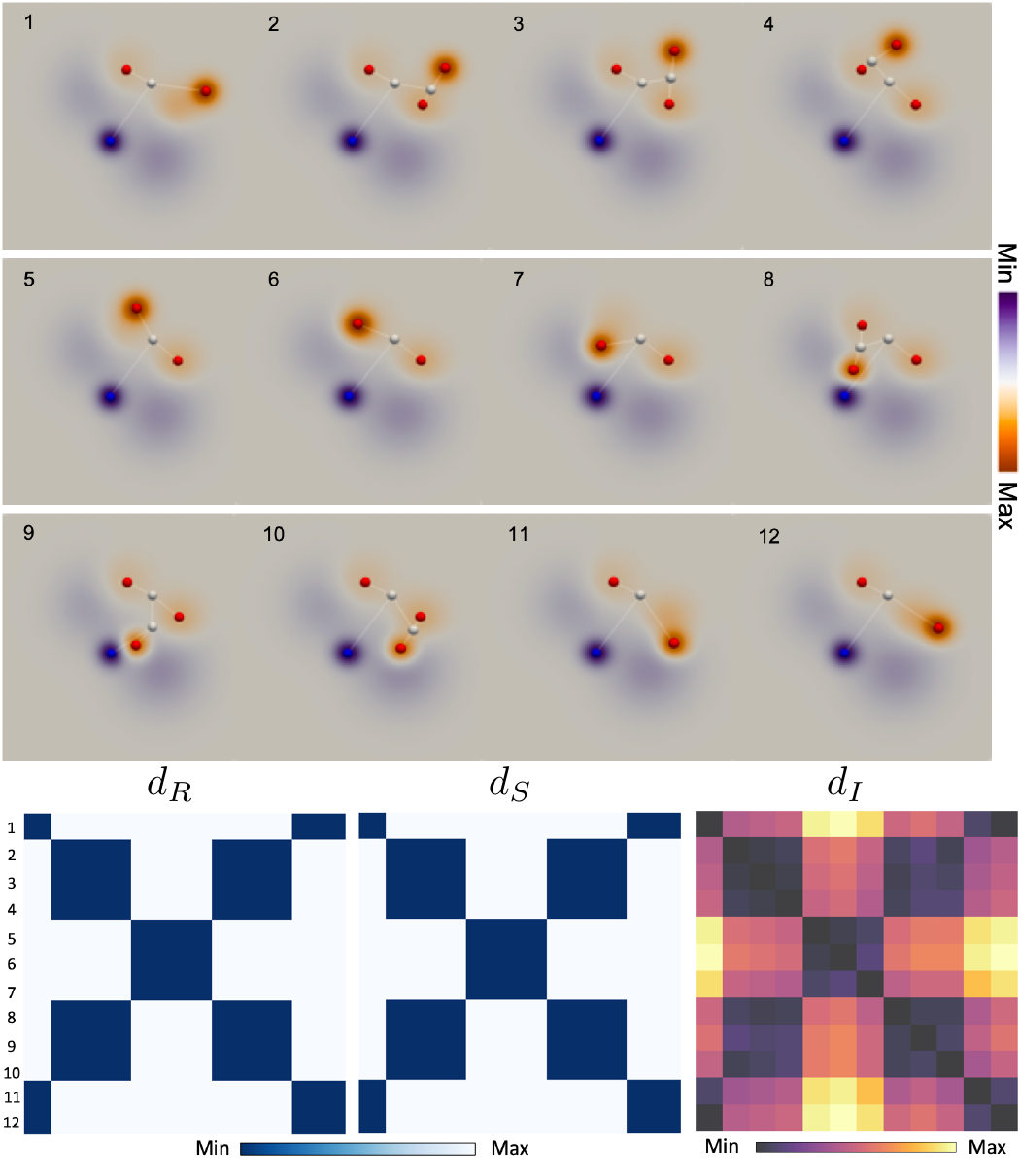}
\vspace{-6mm}    
\caption{Moving Gaussian dataset. Top: 12 time steps are visualized with embedded merge trees. Bottom: comparing merge trees with binary matrices for $d_R$ and $d_S$, together with an interleaving distance $d_I$ matrix.}
\label{fig:Gaussian}
\vspace{-3mm}
\end{figure}

We first consider a toy dataset, called {Moving Gaussian}, generated by placing a mixture of three Gaussian functions on a plane. 
The dataset consists of 12 time steps, where one of the three Gaussian functions moves counterclockwise around two fixed Gaussian functions~\cite{YanMasoodRasheed2023}. 
We compute the merge trees of the inverse, which capture the relationships between local extrema and saddles.  
\cref{fig:Gaussian} (top) illustrate the set of scalar fields along with the merge trees. 

This dataset contains natural clusters by design. 
Yan \etal~\cite{YanMasoodRasheed2023} reported  three clusters formed by time steps $\{2,3,4,8,9,10\}$, $\{5,6,7\}$, and $\{1,11,12\}$, respectively; see~\cref{fig:Gaussian} (bottom right). 

We compare our LSH framework against the results generated with the interleaving distance  in~\cite{YanMasoodRasheed2023}. 
We apply a persistence threshold $\varepsilon=0.02$ to separate features from noise in the scalar fields. 
We then generate merge trees and labels for all 12 time steps and apply our LSH-based similarity measures. 

To visualize our results, we mark an entry in a $12 \times 12$ binary matrix to be $1$ if there exists a candidate pair by an LSH collision, otherwise $0$. 
The binary matrices for both $d_R$ and $d_S$ are shown in~\cref{fig:Gaussian} (bottom), along with a $12 \times 12$ interleaving  distance $d_I$ matrix from~\cite{YanMasoodRasheed2023}.
We observe similar clustering results from both $d_R$ and $d_S$, confirming that our framework can be a good alternative to the interleaving distance. 

We experiment with different values of $k$, $b$,  and $r$. 
Our results match exactly the results  in~\cite{YanMasoodRasheed2023}, when $k=4$, $r=4$, and $b=1$ for $d_R$ and $k=8$, $r=4$, and $b=2$ for $d_S$. 
For subsequent experiments,~\cref{table:data} shows all datasets and parameters used in the experiments. 

\subsection{Shape Matching}
\label{sec:shape}

Shape matching involves detecting similarities between shapes. 
Even though geometric methods have been very powerful in shape matching, we could still use topological descriptors to get reasonable results.
 
\begin{figure}[!ht]
\centering
\vspace{-3mm}
\includegraphics[width=1.0\columnwidth]{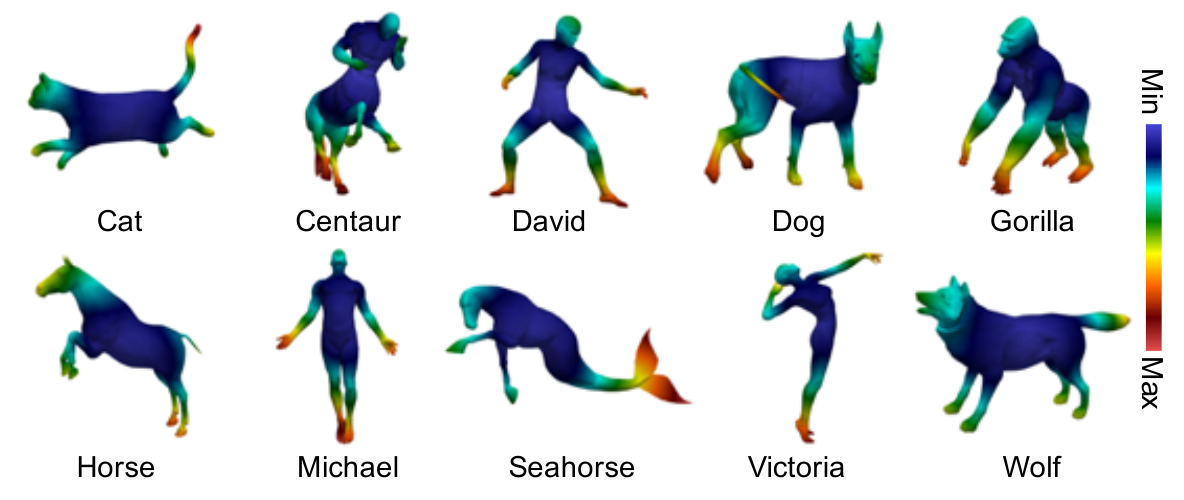}
\vspace{-6mm}    
\caption{TOSCA shape dataset contains 10 classes of shapes. One pose is selected to represent each class. Each mesh is colored by the average geodesic distance from a set of anchor points.}
\label{fig:shapes}
\vspace{-2mm}
\end{figure}

We use a TOSCA shape dataset, which contains various nonrigid shapes in different poses.~\cref{fig:shapes} shows 10 classes of shapes with ground truth labels, where each class contains multiple poses of the same shape.
We aim to classify similar shapes correctly using merge trees, irrespective of their poses.

\begin{figure}[!ht]
\centering
\vspace{-6mm}
\includegraphics[width=1.0\columnwidth]{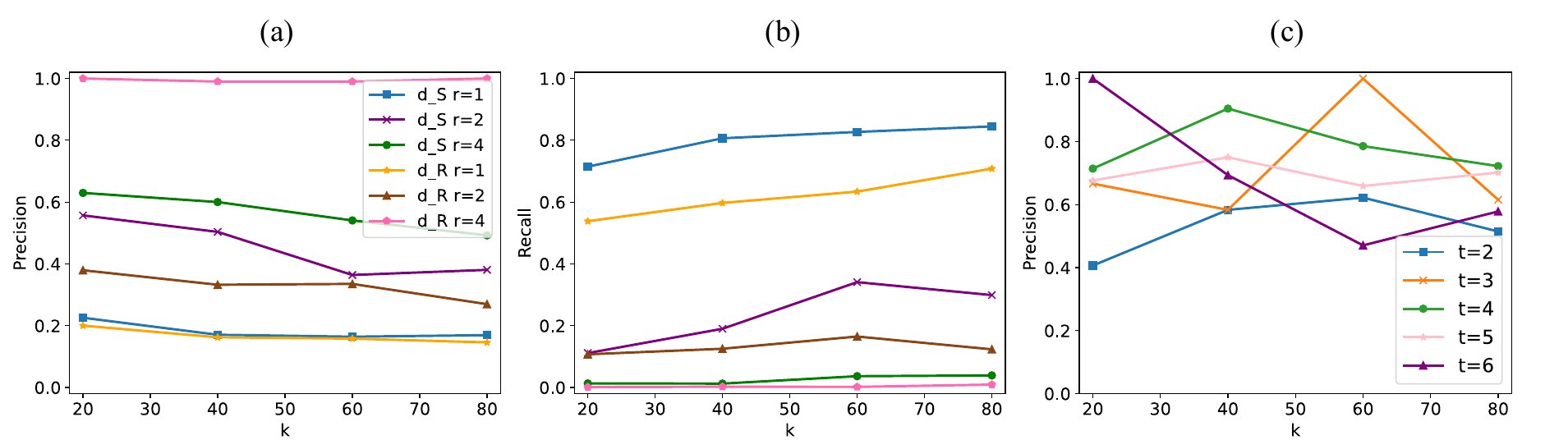}
\vspace{-6mm}    
\caption{TOSCA shape dataset: precision and recall in shape matching. 
(a)-(b): Precision and recall plot for $d_R$ and $d_S$, by  varying $r$ w.r.t $k$.
(c): Exploring parameter configurations for $d_S$.}
\label{fig:shape-precision-recall}
\vspace{-3mm}
\end{figure}
We define the scalar field $f$ to be the average geodesic distance from a set of anchor points on the mesh.  
We then generate merge trees of $-f$ and their node labels.
We use this dataset to demonstrate the utility of our LSH framework in finding exact matching pairs. 
We also use it to illustrate parameter choices.

We first investigate how the choices of parameters $k$ and $r$ (so $b= k/r$) can affect the outcome of $d_R$ and $d_S$ in the context of shape matching where we have labled ground truth.  The same strategy extends to other datasets. 
Note that the primary computational cost of a LSH algorithm is computing the hash function for each data object, where $k$ indeed can be used as a representation of runtime. In \cref{fig:runtime-trend}, we explicitly investigate the relationship between $k$ and runtime.

\cref{fig:shape-precision-recall} plots the average precision and recall as we vary $r$ and $k$.
For each shape $x$, a \emph{precision} is computed as a ratio of the matched instances with the same label over all matched instances.     
A pointwise \emph{recall} is computed as a ratio between the matched instances with the same label versus all instances with that label.  
As $r$ increases, precision increases, but recall decreases (as with any distance and similarity threshold).  We see this for $d_R$ and $d_S$ in this plot considering $k \in [20,40,60,80]$ and $r \in [1,2,4]$.  
We observe $k$ does not affect precision and recall much and so we often opt for a smaller $k=20$.  
We also favor a smaller $r=1$, since this gives much better recall.  We can always filter mis-matches (because of not as good precision) with a direct computation of a slower distance over a smaller number of matched objects.  

\begin{figure}[!ht]
\centering
\vspace{-4mm}
    \includegraphics[width=0.8\columnwidth]
    {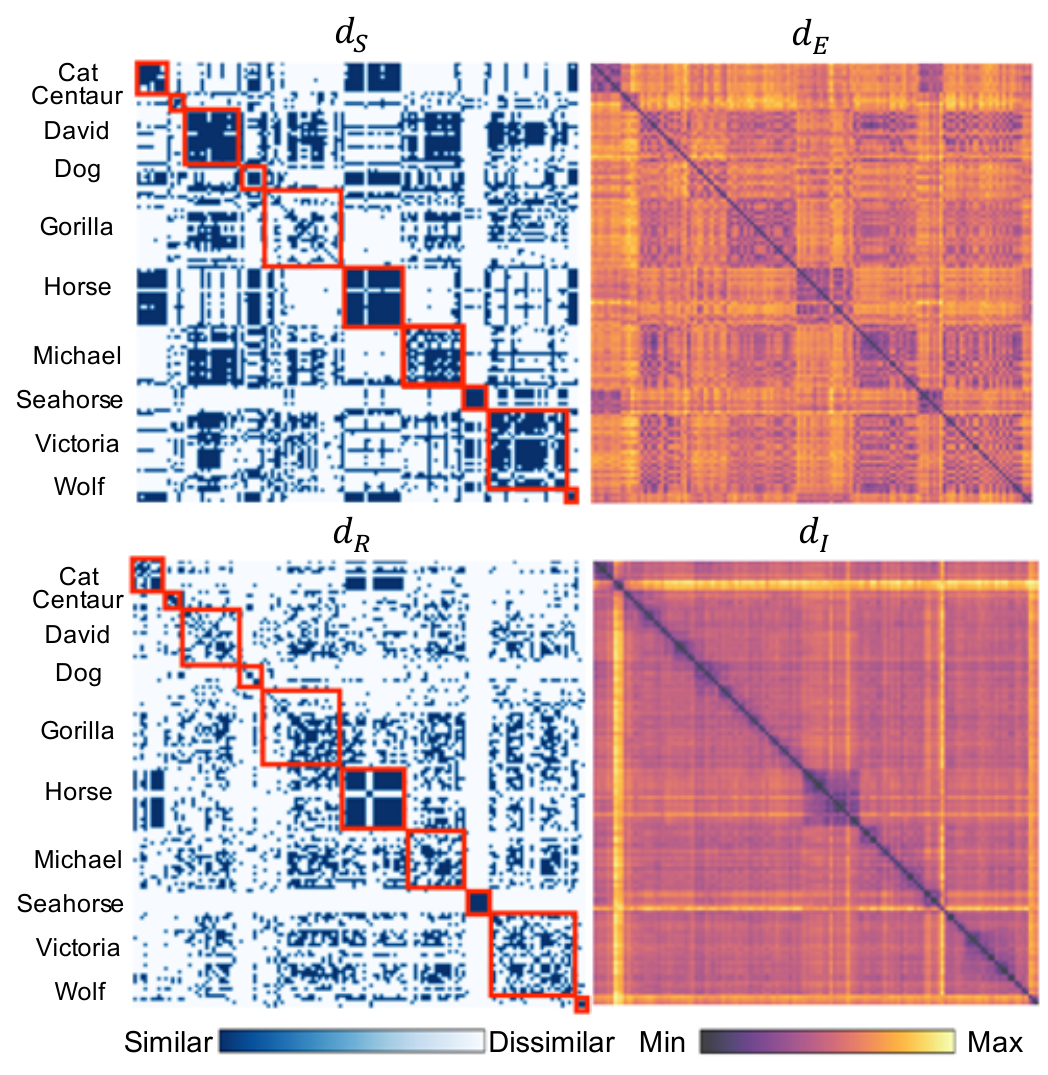}
\vspace{-4mm}    
\caption{TOSCA shape dataset.  
Binary matrices of $d_R$ and $d_S$ are shown with red squares which contain instances with the same ground truth class label. 
$d_E$ and $d_I$ distance matrices are provided for context.}
\label{fig:shape-matrix}
\vspace{-3mm}
\end{figure}

\cref{fig:shape-precision-recall} also varies subpath length $t$.  We choose $t=4$ to maintain a good precision without the subpath being too long, depending on the height of the merge trees. 

We perform experiments on different parameter settings provided in~\cref{table:data}. 
\cref{fig:shape-matrix} shows the binary matrices for both $d_R$ and $d_S$, with $k = 20$ and $r = 1$. 
Each binary matrix shows clear block structures along its diagonal, for example, in Cat, David, Horse, Michael, Seahorse, and Victoria classes. 
The results show that we obtain reasonable classification of some classes. 

On the other hand, $d_R$ and $d_S$ (to some extent) perform imperfect classification between Gorilla, Horse, and Michael classes. For instance, we observe off diagonal blocks showing similarities between Horse and Cat.  
The reason is two-fold. 
First, merge trees have similar structures among these classes of shapes, and the merge tree itself is not always a good descriptor for capturing all the geometric details of a shape. 
Second, the labeling strategy, which currently labels only the leaf nodes, might also contribute toward the imperfect results. 
The LSH framework intuitively captures a variation of Jaccard similarity (based on hash buckets). Since the Jaccard similarity generally ignores hierarchy, $d_R$ applies Jaccard similarity at multiple levels to alleviate this issue. 

In~\cref{fig:shape-matrix}, $d_E$ and $d_I$ matrices are also provided for context. 
In comparison, $d_I$ captures less information as fewer classes are classified correctly by $d_I$.  
We observe that $d_E$ gives good classification results for only some of the classes. For instance, $d_E$ also exhibit off diagonal blocks showing similarities between Gorilla and Michael, David and Michael, and so on.
The small discrepancies between $d_E$ and our framework probably arise from modeling shapes with merge trees, not the similarity measures we employ. 

\vspace{-1mm}
\subsection{Time-Varying and Ensemble Data Summarization}
\label{sec:time-varying}

We demonstrate via experiments the utility of our LSH framework in summarizing time-varying scalar fields and ensembles. 

\subsubsection{Vortex Street}
\label{sec:vortex-street}
\begin{figure}
\centering
\includegraphics[width=0.5\textwidth]
	{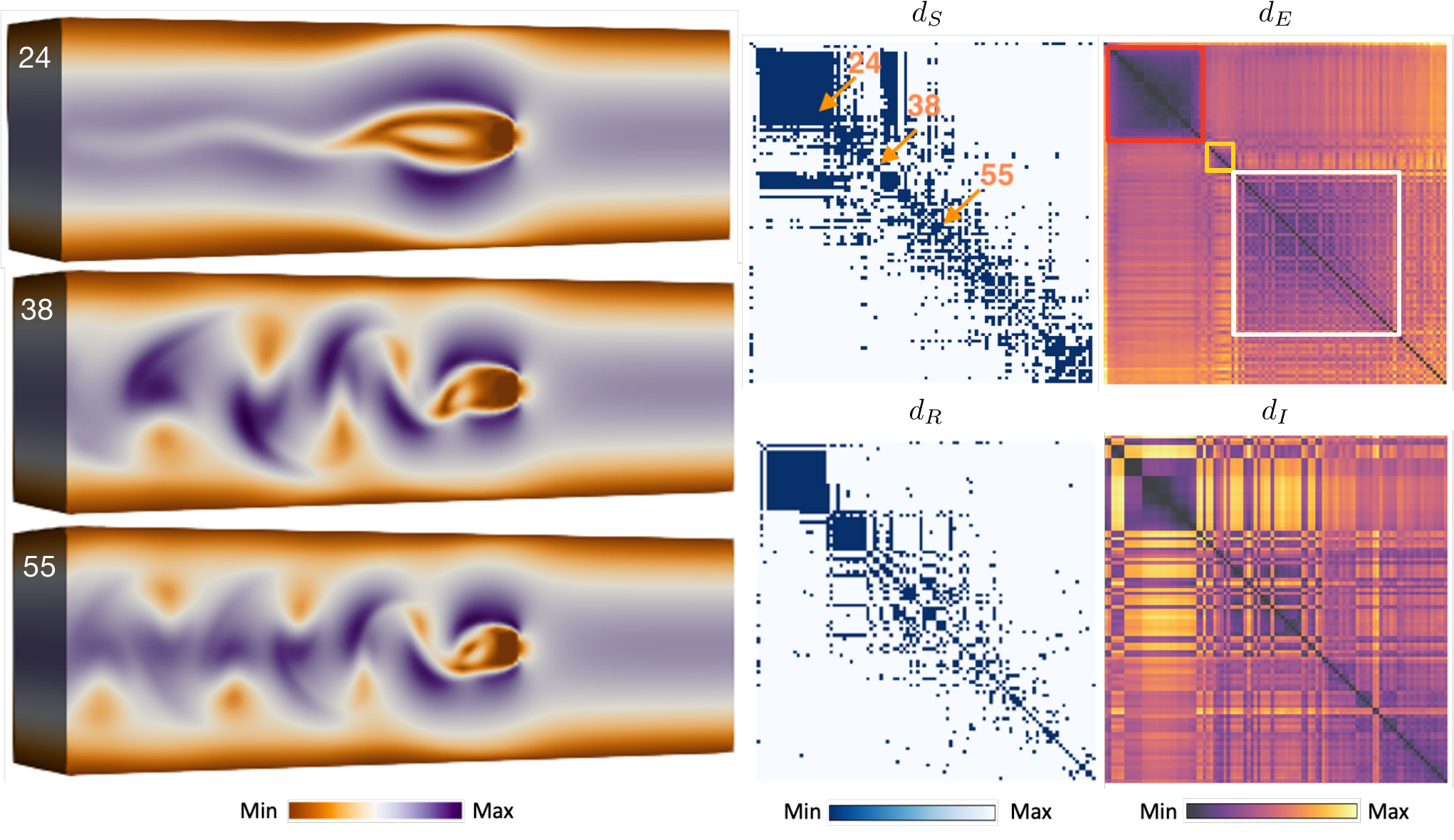}
\vspace{-6mm}	
\caption{Vortex Street dataset. Binary matrices for $d_R$ and $d_S$ are presented with selected time steps from different clusters. Distance matrices for $d_I$ and $d_E $ are provided for comparison.}
\label{fig:vortex-street}
\vspace{-6mm}
\end{figure}

We first demonstrate temporal summarization using a 3D B\'enard von K\'arman vortex street dataset. 
We consider the magnitude of velocity as the scalar field. 
We generate the binary matrices according to the parameters provide in~\cref{table:data}.

\cref{fig:vortex-street} (left) visualizes three instances with different structures along the time-varying dataset. 
This dataset contains a set of clusters capturing different phases of the flow behavior~\cite{SridharamurthyMasoodKamakshidasan2020}. 
During the 1st phase, the flow evolves slowly, and the first 30 time steps are highly similar to one another, and time step 24 is selected as a representative. 
During the 2nd phase, the flow is transitioning toward a more periodic behavior, and time step 38 marks a transition to the next phase.  
During the 3rd phase, the vortex shedding phenomena (i.e., an oscillating flow behind a bluff body at certain velocities) become clearly visible, and time step 55 is selected as a representative. 

\cref{fig:vortex-street} (right) includes binary matrices $d_R$ ($k = 60$, $r = 1$) and $d_S$ ($k = 20$, $r = 1$, $t=4$).  
We observe that $d_E$ clearly identifies the three phases as three clusters (blocks). In particular, it shows periodicity in the 3rd block that corresponds to vortex shedding.  
In the $d_S$ matrix, we identify the 1st major cluster clearly, but the 2nd minor cluster and the 3rd cluster are not as obvious as those from $d_I$. 
In the $d_R$ matrix, the vortex shedding is vaguely visible as periodic block patterns, similar to what is shown in the $d_I$ matrix.  

\subsubsection{Viscous Finger} 
\label{sec:viscous-finger}

We use a Viscous Finger dataset to demonstrate the scalability of our LSH method in summarizing a large-scale ensemble dataset and uncover similarity between particular runs. 

The ensemble is composed of 3D transient fluid flow obtained by a simulation with stochastic effects, formulating a special behavior named as viscous finger. 
We choose the ensemble with $0.44$ as the resolution level. 
This ensemble contains 48 runs, each contains $\approx 120$ time steps. To compare equally among different runs, we generate binary matrices across all runs with parameters in~\cref{table:data}.

\begin{figure}
\centering
\includegraphics[width=0.7\columnwidth]{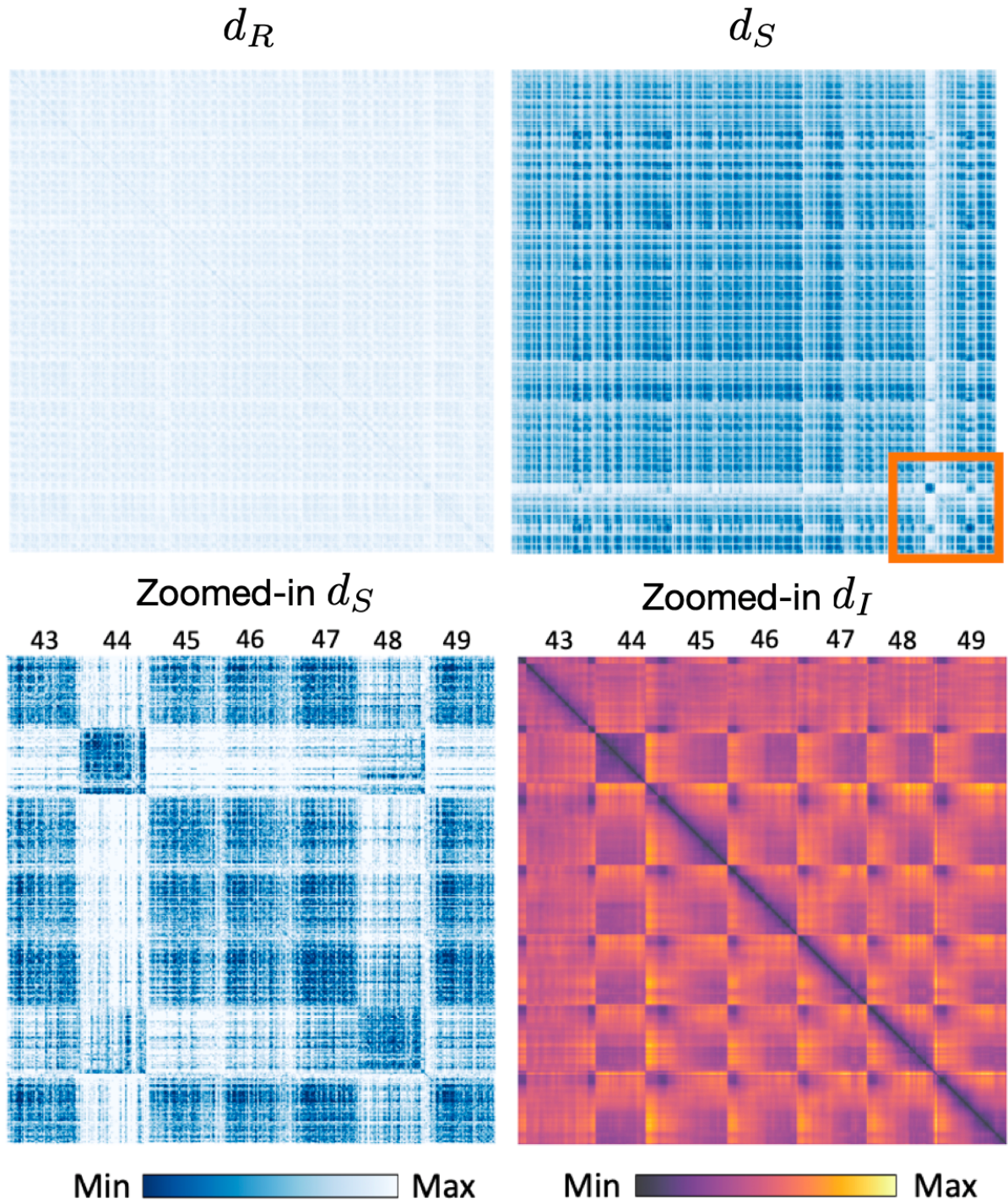}
\vspace{-2mm}
\caption{Viscous Finger dataset. Top: binary matrices for $d_R$ and $d_S$. Bottom: the zoom-in matrices are from run 43 to run 49.}
\label{fig:viscous-finger}
\end{figure}

\cref{fig:viscous-finger} (top) shows binary matrices $d_S$ (with $k=60$ and $r=1$) and $d_R$ ($k=20$ and $r=1$) across 48 runs.  
For this large dataset, both $d_E$ and $d_I$ become intractable; therefore, we compute only a submatrix of $d_I$ for comparison. 
The binary matrix $d_R$ has a lighter color compared to $d_S$ as $d_R$ shows less similarities than $d_S$. 

We observe that most runs are similar due to a similar simulation process, except the end instance and the beginning instance connecting two runs where straight line patterns separate the entire matrix into blocks. 
From the zoomed-in $d_I$ matrix, we also observe such block-wise pattern. 

\begin{figure}
\centering
\vspace{-4mm}
\includegraphics[width=1.0\columnwidth]{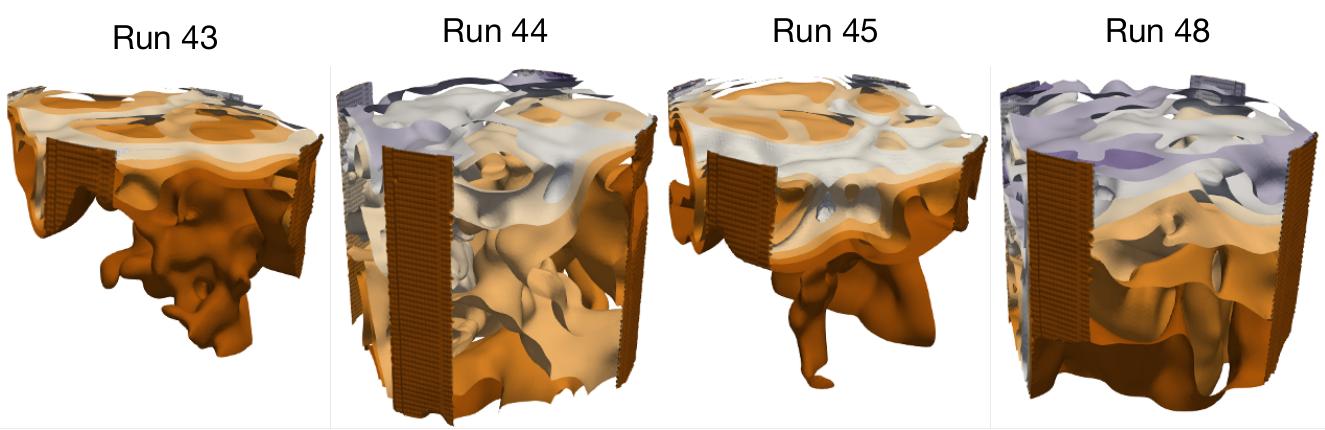}
\vspace{-6mm}
\caption{Viscous Finger dataset. Isosurface rendering of time step 120 from runs 43, 44, 45, and 48, respectively.}
\label{fig:viscous-finger-run}
\vspace{-6mm}
\end{figure}

\begin{figure}[!ht]
\centering
\includegraphics[width=0.8\columnwidth]{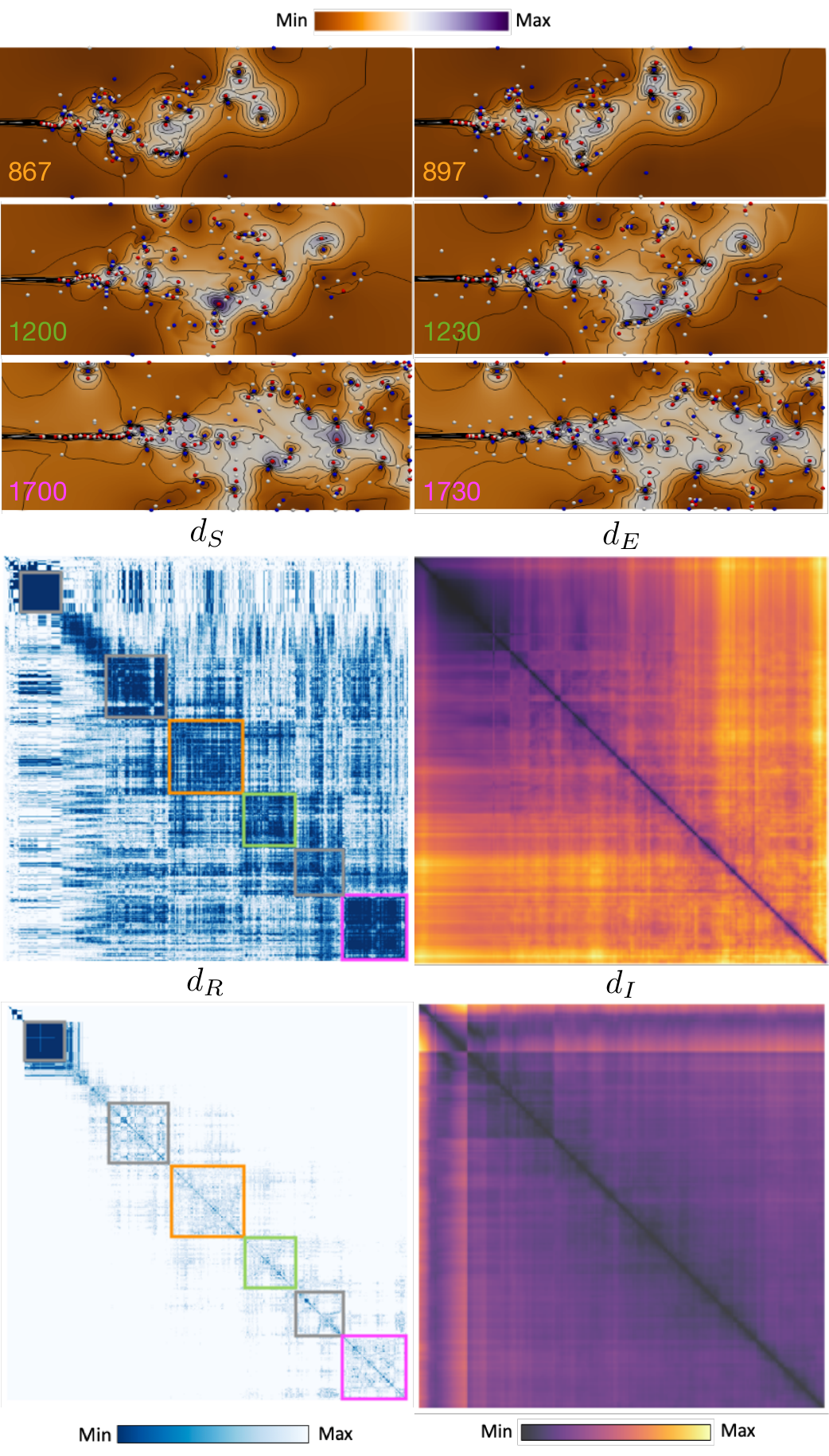}
\vspace{-2mm}
\caption{Heated Flow dataset. Top: time steps from different clusters. Bottom: $d_R$ and $d_S$. Different colored boxes represent different clusters.}
\label{fig:heated-flow}
\vspace{-6mm}
\end{figure}

Our binary matrices identify two runs that perform differently, runs 44 and 48. 
We also observe similar patterns in the $d_I$ matrix,  where run 44 forms a clear diagonal block, meaning that instances inside the run are more similar internally than externally. 

We select time step 120 from each of the runs 43, 44, 45, and 48, and render their isosurfaces respectively, shown in~\cref{fig:viscous-finger-run}. 
We observe that runs 44 and 48 behave similarly with each other. 
These visualizations match the observations from the binary matrices in~\cref{fig:viscous-finger}. 

Given a large ensemble, our LSH framework manages to capture the global behavior due to its efficiency and scalability, whereas $d_E$ and $d_I$ fail to do so. 
In particular, both $d_S$ and $d_R$ allow us to discover similarities among the runs.

\subsection{Identification of Clusters and Data Transitions}
\label{sec:clusterings}

Here we show the utility of our framework in identifying clusters and structural transitions in large flow datasets. 

\subsubsection{Heated Flow Dataset}
\label{sec:heated-flow}

The Heated Flow dataset is a time-varying 2D dataset generated by flow around a heated cylinder using Boussinesq  approximation where the flow contains many small vortices. 
We use the magnitude of the flow as the scalar field $f$. 
We compute merge trees of $f$ and node labels. 
The binary matrices are generated based on the parameters in~\cref{table:data}.

\cref{fig:heated-flow} shows binary matrices of $d_S$ ($k=60$, $r=1$) and $d_R$ ($k=20$ and $r=1$).
Our framework helps to identify six clusters, highlighted as colored blocks: cluster 1 (grey), time steps 70 - 300; cluster 2 (grey), time steps 485 - 814; cluster 3 (orange), time steps 815 - 1180; cluster 4 (green), time steps 1181 - 1450; cluster 5 (grey), time steps 1451 - 1690; and cluster 6 (magenta), time steps  1691 - 2000. 

We select two time steps, 867 and 897, 1200 and 1230, and 1700 and 1730, from each of the three clusters (orange, green, and magenta) to show their similarity within the cluster and dissimilarity outside the cluster. 
The colors on the labels correspond to different clusters. 

$d_E$ matrix visibly captures only the 1st cluster, but fails to show noticeable clusters for the rest of the dataset. 
$d_I$ matrix, on the other hand, groups most time steps into one big cluster.
Therefore, our LSH framework provides clearer clustering structure. 

\subsubsection{Corner Flow Dataset}
\label{sec:corner-flow}

The Corner Flow is a 2D dataset generated by flow around two cylinders. 
The flow is bound by walls with corners around which vortices form due to the presence of the cylinders.
We use the velocity magnitude as the scalar field $f$.
We generate the binary matrices using parameters  in~\cref{table:data}. 
We demonstrate the ability of our framework in capturing structural transitions of time-varying scalar fields. 

\begin{figure}[!ht]
\centering   
\includegraphics[width=0.87\columnwidth]{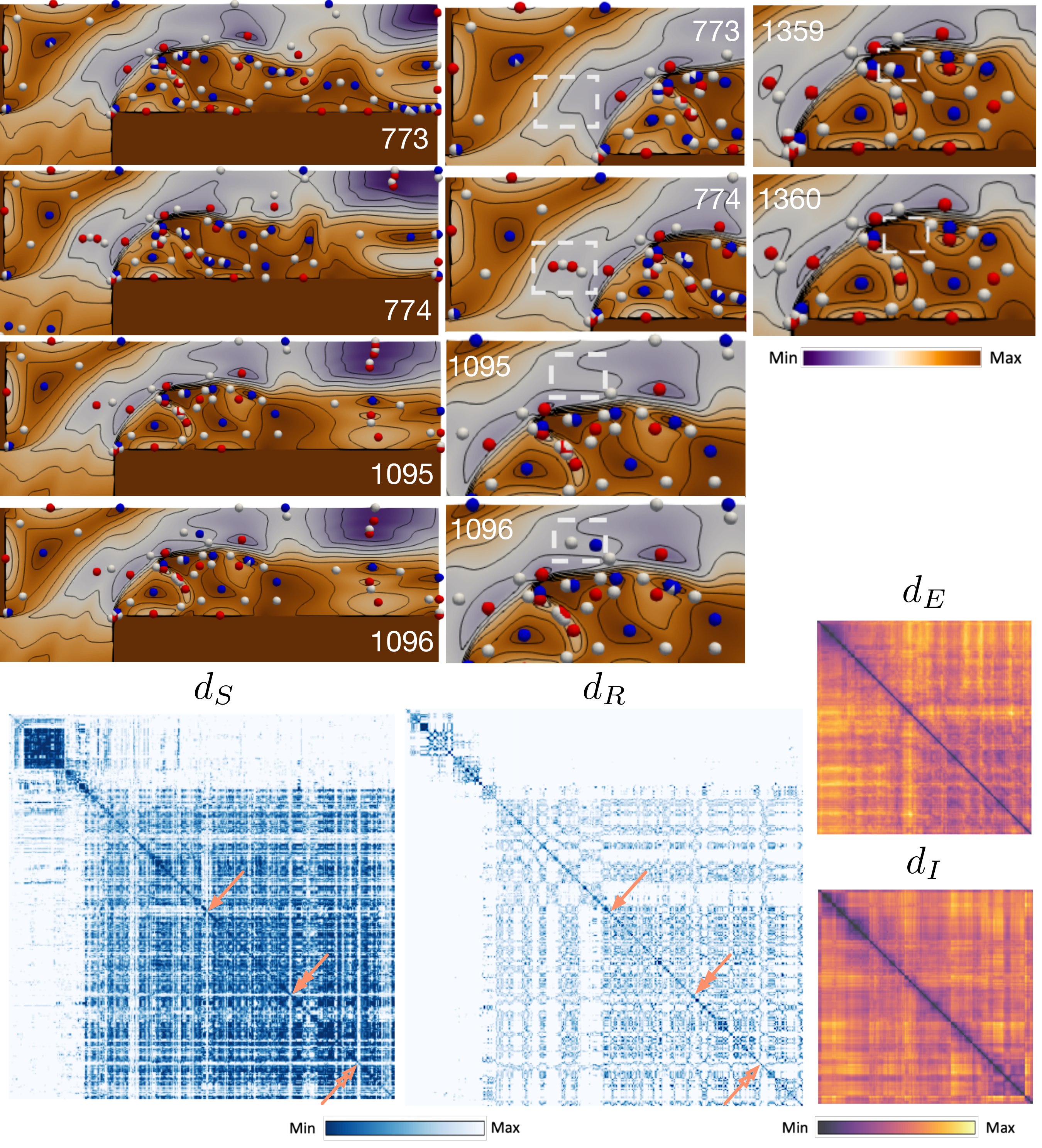}
\vspace{-4mm}    
\caption{Corner Flow dataset. $d_S$ and $d_R$: binary matrices with visualized scalar fields to show structural transitions. $d_E$ and $d_I$ are also provided.}
\label{fig:corner-flow}
\end{figure}

\cref{fig:corner-flow} shows the binary matrices of $d_R$ ($k = 20$, $r = 1$) and $d_S$ ($t=4$). 
We observe three pairs of time steps showing structural transitions, highlighted as arrows in  $d_S$ and $d_R$. 
From time step 773 to 774 (orange filled arrow), there is a structural transition. As fluid flows, the contour line expands and two max-saddle pairs disappear in 774. 
The critical points are highlighted in the white dashed box. 
From time step 1095 to 1096 (orange filled double arrow), there is another structural transition where a new min-saddle pair appears. 
From time step 1359 to 1360 (orange double arrow), a min-saddle pair disappears. 

\begin{table}[]
\scriptsize
\vspace{-6mm}
\caption{Runtime analysis of $d_S$ and $d_R$. Each entry represents the LSH runtime with a corresponding $k$. All times are in seconds.}
\vspace{-2mm}
\begin{tabular}{l|c|c|c|c|c}
\hline
\begin{tabular}[c]{@{}l@{}}Dataset\\ Labeling time\end{tabular}           & Method & $k=20$ & $k=40$ & $k=60$ & $k=80$ \\ \hline
\multirow{2}{*}{\begin{tabular}[c]{@{}l@{}}Vortex Street\\ \textbf{37.9}\end{tabular}}      & $d_S$  & 0.18   & 0.21   & 0.23   & 0.27   \\  
& $d_R$ & 1.26   & 4.19   & 10.55  & 21.82  \\ \hline
\multirow{2}{*}{\begin{tabular}[c]{@{}l@{}}TOSCA\\ \textbf{11.45}\end{tabular}}             & $d_S$  & 0.27   & 0.30   & 0.34   & 0.38   \\  
 & $d_R$ & 1.28   & 4.03   & 10.07  & 20.77  \\ \hline
\multirow{2}{*}{\begin{tabular}[c]{@{}l@{}}Corner Flow\\ \textbf{360.07}\end{tabular}}      & $d_S$  & 3.85   & 4.51   & 5.09   & 5.71   \\  
& $d_R$ & 37.49  & 121.17 & 307.22 & 637.55 \\ \hline
\multirow{2}{*}{\begin{tabular}[c]{@{}l@{}}Heated Flow\\ \textbf{2659.2}\end{tabular}}      & $d_S$  & 1.5    & 1.72   & 2.03   & 2.31   \\ 
& $d_R$  & 76.9   & 254    & 663.2  & 1357.2 \\ \hline
\multirow{2}{*}{\begin{tabular}[c]{@{}l@{}}Viscous Fingers\\ \textbf{13779.9}\end{tabular}} & $d_S$  & 45.1   & 55.18  & 65.45  & 72.18  \\ 
& $d_R$  & 374    & 1156   & 2843   & 5636   \\ \hline
\end{tabular}
\label{table:runtime1}
\end{table}

\begin{table}[]
\scriptsize
\vspace{-4mm}
\caption{Runtime analysis across datasets: columns $d_S$ and $d_R$ represent the total runtime (the sum of LSH time and labeling time) for $k=20$. The improved runtime are highlighted in bold. All times are in seconds. Numbers with asterisk are estimated.}
\vspace{-2mm}
\begin{tabular}{c|c|c|c|c}
\hline
Dataset         & $d_S$     & $d_R$     & $d_I$       & $d_E$         \\
\hline
Vortex Street   & \textbf{38.08}  & \textbf{39.26}  & 65.5     & 48.0       \\ 
TOSCA           & \textbf{11.72}  & \textbf{12.72}  & 24.9     & 109.7      \\ 
Corner Flow     & \textbf{363.92} & \textbf{397.56} & 3456.0   & 6203.1     \\ 
Heated Flow     & \textbf{2660.7} & \textbf{2736.1} & 23686    & 211688     \\
Viscous Fingers & \textbf{13825}  & \textbf{14154}  & $\approx$455388$^{*}$ & $\approx$15965520$^{*}$ \\ \hline
\end{tabular}
\label{table:runtime2}
\vspace{-4mm}
\end{table}

$d_R$ matrix shows similar patterns as $d_E$ matrix,  where a small cluster connects to a large cluster for the 2nd half of the time-varying dataset. 
$d_R$ matrix also presents similar patterns (i.e., two clusters) as the $d_I$ matrix at the beginning of the time steps. 
The binary matrices of both $d_R$ and $d_S$ can capture transitions, whereas $d_S$ provides a more precise structural detection. 
In conclusion, our LSH framework serves as a good alternative to existing distance measures.

\subsection{Quantification of Runtime Improvements}
\label{sec:runtime}

\cref{table:runtime1} and \cref{table:runtime2} provide a runtime analysis of our LSH framework, which depends on two main factors. 
The first is the creation of $k$ signatures, and the use of hash buckets to identify matchings. 
The second is the labeling of the merge trees following a hybrid strategy in~\cite{YanMasoodRasheed2023}. 
\cref{table:runtime1} reports LSH runtime, for both $d_S$ and $d_R$ along with the labeling time in bold. For both $d_S$ and $d_R$, the LSH runtime increases as $k$ increases. 
We also observe that $d_S$ is faster than $d_R$ for all datasets.

From~\cref{table:runtime1}, we can see that as the size of the data grows, the LSH process has a very small runtime, and the runtime bottleneck is indeed the labeling process.   
We argue that this is acceptable in certain scenarios. For instance, assuming we are performing interactive analysis in real time over a fixed dataset, the labeling process can be precomputed (once), and only the LSH process needs to run multiple times on-the-fly. 

For comparison, we report runtime for both $d_E$ and $d_I$ along with the total runtime for $d_S$ and $d_R$ for $k=20$ in \cref{table:runtime2}. The total runtime is the sum of LSH time and the labeling time.
For the Vortex Street dataset, we have a similar runtime for most cases compared to $d_E$ and $d_I$, where $d_R$ gives slightly higher runtime when $k=60$ and $k=80$. 
For the TOSCA dataset, $d_R$ gets slightly higher runtime compared to $d_I$, when $k=80$; otherwise our LSH measures are faster to compute. 
We obtain approximately $3\times$ speed-up for $d_R$ ($k=80$) compared to $d_E$. 
Our LSH framework has approximately $1.5\times$ speed-up compared to $d_I$, and approximately  $9\times$ speed-up compared to $d_E$ for $d_S$. Our LSH framework takes less time for TOSCA than Vortex Street. Despite TOSCA containing more instances, the merge trees are generally smaller. 

We observe a larger speedup for Corner Flow and Heated Flow. 
For both datasets, we have (approximately) $9\times$ speedup for $d_S$ and $6\times$ speed-up for $d_R$ compared to $d_I$. 
For Corner Flow, we have $17\times$ and $6\times$ speed-up compared to $d_E$ for $d_S$ and $d_R$, respectively. 
For Heated Flow, we have $80\times$ for $d_S$ and $5\times$ speed-up for $d_R$ compared to $d_E$.

For Viscous Finger, the runtime is significantly improved. 
We compute $d_I$ for only a subset of the data, since computing the entire $d_I$ matrix is estimated to take $5$ days; this runtime is estimated based on the recorded runtime for the first $40,000$ comparisons. 

We obtain $33\times$ ($d_S$) and $23\times$ ($d_R$) speed-up compared to $d_I$, even for $k=80$. 
We are not able to compute $d_E$ for the full dataset comprising of $5746$ instances, due to its lack of scalability in computation.  
Therefore, we run $d_E$ on one single run that contains $120$ instances, and report a runtime of $\approx 6929$ seconds, close to 2 hours. 
We estimate the runtime of $d_E$ for the entire dataset, estimating across all $48$ runs, which takes roughly $48^2 \times 6929 \approx 185$ days! 
Thus, our LSH approach is estimated to be about $822 \times$ faster than $d_E$ computation. 
  
As data grow larger, we observe larger  improvements, compared to $d_E$ and $d_I$. 
$d_E$ suffers from poor scalability due to two reasons. First, all pairs comparisons need to be  explicitly performed. Second, each comparison involves solving a set of matching problems, which leads to high runtime in practice. 
Even for the Viscous Finger dataset, $d_E$ and $d_I$ distance matrices become impractical to compute, and our LSH framework becomes a necessity. 

\begin{figure}[!ht]
\centering
\vspace{-4mm}
\includegraphics[width=1.0\columnwidth]
    {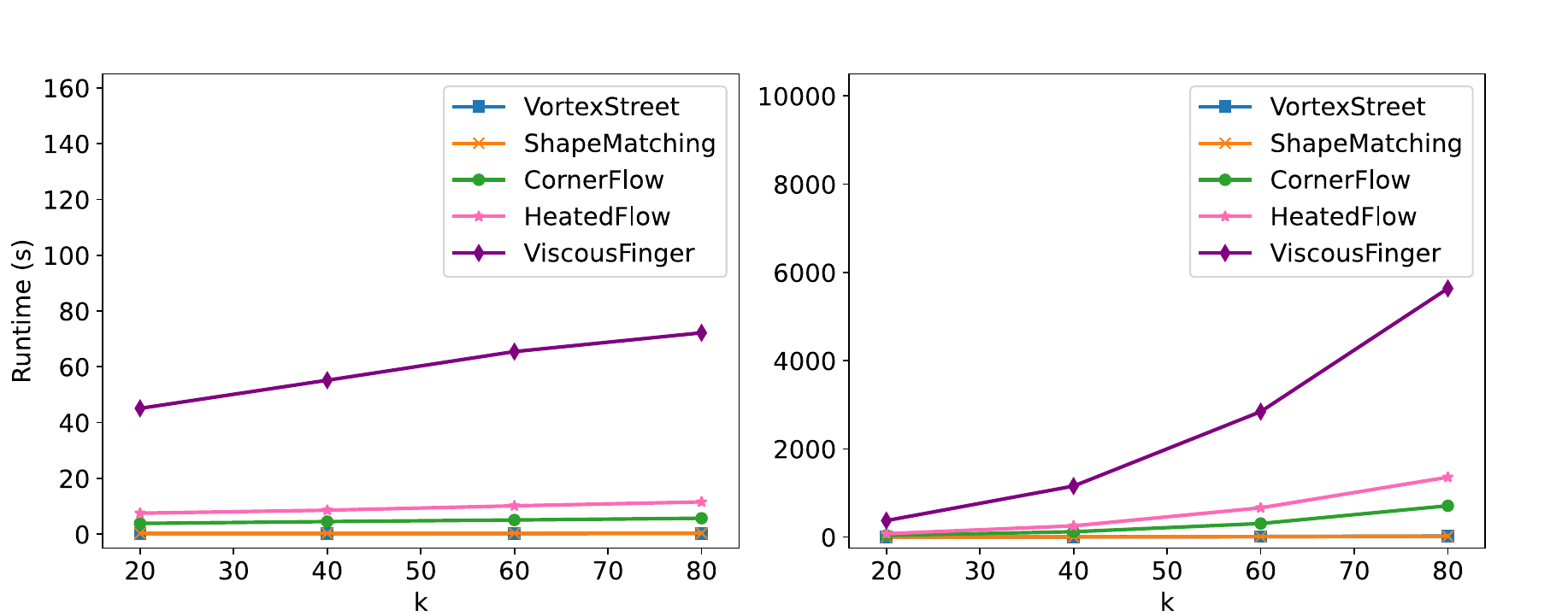}
\vspace{-6mm}
\caption{Runtime analysis of $d_S$ and $d_R$, excluding the labeling process. 
Left: $k$ vs. runtime for $d_S$, $t=4$. 
Right: $k$ vs. runtime for $d_R$.}
\label{fig:runtime-trend}
\vspace{-4.5mm}
\end{figure}

In addition, we perform an ablation runtime analysis as we vary $k$ in~\cref{fig:runtime-trend}.  
Our analysis shows the LSH runtime excluding the labeling process, with different $k$.   
$d_S$ presents nearly linear behavior and $d_R$ shows nearly quadratic behavior as a function of $k$. There are other costs associated with retrieving the query, but they are lower-order terms and do not depend on the size of the data object. \cref{fig:runtime-trend} demonstrates that $k$ is a good representation of runtime.
 
\section{Discussion and Limitations}
\label{sec:conclusion}

Our LSH framework using $d_S$ and $d_R$, as expected, is efficient and scalable, in comparison with existing distance methods such as $d_E$ and $d_I$. 
In fact, our framework is efficient on large datasets when $d_E$ and $d_I$ become intractable to compute. 
Furthermore, the $d_S$ and $d_R$ we propose offer enormous advantages in scalability, often replicate results of $d_E$ and $d_I$, and sometimes capture new cluster features that $d_E$ and $d_I$ do not.  
Therefore, they could be used as efficient, alternative measures for comparing scalar fields at scale. As a preprocessing step, LSH can eliminate unnecessary candidates for comparative analysis, therefore reducing the size of input data for other distance computations. 

A limitation of our LSH framework is that it does not recover the nearest-neighbor exactly as does $d_E$. With any LSH approach, we need to adjust parameters $k$ and $r$, which control the trade-off between precision and recall. We find that even small values $k=20, r=1$ provide decent results, and recommend these parameters for similar problems to the ones we consider. 
Another factor affecting LSH is the labeling strategy. While the requirement for labeled merge tree seems like a limitation, the labels can be used to incorporate geometric information which is crucial in many applications~\cite{YanMasoodRasheed2023, CarrSnoeyinkVanDePanne2010, SolerPlainchaultConche2018}.  
We conjecture that our LSH results would improve with more stable labeling strategies for internal nodes. 
Making the labeling process more efficient will also benefit the overall runtime, which is left for future work.

\section*{Acknowledgement}
This work was supported by grants from DOE~DE-SC0021015 and~DE-SC0023157, as well as NSF~DMS-2134223,~IIS-1910733,~IIS-1816149,~CCF-2115677, and~IIS-2311954. 

\bibliographystyle{abbrv-doi-hyperref}
\bibliography{refs-LSH}

\begin{thebibliography}{10}

\bibitem{3Dvs2024}
{3D} flow around a confined square cylinder.
\newblock \url{http://tinoweinkauf.net/notes/squarecylinder.html}, 2024.

\bibitem{AcharyaNatarajan2015}
A.~Acharya and V.~Natarajan.
\newblock A parallel and memory efficient algorithm for constructing the contour tree.
\newblock In {\em IEEE Pacific Visualization Symposium}, pp. 271--278. IEEE, 2015. \href{https://doi.org/10.1109/PACIFICVIS.2015.7156387}
{doi: {{%
10\hspace{.1pt}\discretionary{.}{%
}{.}\hspace{.4pt}1109\discretionary{/}{%
}{/}PACIFICVIS\hspace{.1pt}\discretionary{.}{%
}{.}\hspace{.4pt}2015\hspace{.1pt}\discretionary{.}{%
}{.}\hspace{.4pt}7156387}}}


\bibitem{AiolliDaSanMartinoSperduti2007}
F.~Aiolli, G.~Da~San~Martino, A.~Sperduti, and A.~Moschitti.
\newblock Efficient kernel-based learning for trees.
\newblock In {\em IEEE Symposium on Computational Intelligence and Data Mining}, pp. 308--315. IEEE, 2007. \href{https://doi.org/10.1109/CIDM.2007.368889}
{doi: {{%
10\hspace{.1pt}\discretionary{.}{%
}{.}\hspace{.4pt}1109\discretionary{/}{%
}{/}CIDM\hspace{.1pt}\discretionary{.}{%
}{.}\hspace{.4pt}2007\hspace{.1pt}\discretionary{.}{%
}{.}\hspace{.4pt}368889}}}


\bibitem{Baeza-RojoGunther2020}
I.~Baeza~Rojo and T.~G{\"u}nther.
\newblock Vector field topology of time-dependent flows in a steady reference frame.
\newblock {\em IEEE Transactions on Visualization and Computer Graphics}, 26(1):280--290, 2020. \href{https://doi.org/10.1109/TVCG.2019.2934375}
{doi: {{%
10\hspace{.1pt}\discretionary{.}{%
}{.}\hspace{.4pt}1109\discretionary{/}{%
}{/}TVCG\hspace{.1pt}\discretionary{.}{%
}{.}\hspace{.4pt}2019\hspace{.1pt}\discretionary{.}{%
}{.}\hspace{.4pt}2934375}}}


\bibitem{BajajGilletteGoswami2009}
C.~Bajaj, A.~Gillette, and S.~Goswami.
\newblock Topology based selection and curation of level sets.
\newblock In H.-C. Hege, K.~Polthier, and G.~Scheuermann, eds., {\em Topology-Based Methods in Visualization II}, pp. 45--58. Springer, 2009. \href{https://doi.org/10.1007/978-3-540-88606-8_4}
{doi: {{%
10\hspace{.1pt}\discretionary{.}{%
}{.}\hspace{.4pt}1007\discretionary{/}{%
}{/}978\discretionary{%
}{-}{-}3\discretionary{%
}{-}{-}540\discretionary{%
}{-}{-}88606\discretionary{%
}{-}{-}8\_4}}}


\bibitem{BeketayevYeliussizovMorozov2014}
K.~Beketayev, D.~Yeliussizov, D.~Morozov, G.~H. Weber, and B.~Hamann.
\newblock Measuring the distance between merge trees.
\newblock In P.-T. Bremer, I.~Hotz, V.~Pascucci, and R.~Peikert, eds., {\em {Topological Methods in Data Analysis and Visualization III}}, pp. 151--165. Springer, 2014. \href{https://doi.org/10.1007/978-3-319-04099-8_10}
{doi: {{%
10\hspace{.1pt}\discretionary{.}{%
}{.}\hspace{.4pt}1007\discretionary{/}{%
}{/}978\discretionary{%
}{-}{-}3\discretionary{%
}{-}{-}319\discretionary{%
}{-}{-}04099\discretionary{%
}{-}{-}8\_10}}}


\bibitem{BiedertGarth2015}
T.~Biedert and C.~Garth.
\newblock Contour tree depth images for large data visualization.
\newblock In {\em Proc. 15th Eurographics Symposium on Parallel Graphics and Visualization}, pp. 77--86, 2015. \href{https://doi.org/10.2312/pgv.20151158}
{doi: {{%
10\hspace{.1pt}\discretionary{.}{%
}{.}\hspace{.4pt}2312\discretionary{/}{%
}{/}pgv\hspace{.1pt}\discretionary{.}{%
}{.}\hspace{.4pt}20151158}}}


\bibitem{BollenTennakoonLevine2023}
B.~Bollen, P.~Tennakoon, and J.~A. Levine.
\newblock Computing a stable distance on merge trees.
\newblock {\em {IEEE Transactions on Visualization and Computer Graphics}}, 29(01):1168--1177, 2023. \href{https://doi.org/10.1109/TVCG.2022.3209395}
{doi: {{%
10\hspace{.1pt}\discretionary{.}{%
}{.}\hspace{.4pt}1109\discretionary{/}{%
}{/}TVCG\hspace{.1pt}\discretionary{.}{%
}{.}\hspace{.4pt}2022\hspace{.1pt}\discretionary{.}{%
}{.}\hspace{.4pt}3209395}}}


\bibitem{Broder2000}
A.~Z. Broder.
\newblock Identifying and filtering near-duplicate documents.
\newblock In R.~Giancarlo and D.~Sankoff, eds., {\em Combinatorial Pattern Matching}, pp. 1--10. Springer, 2000. \href{https://doi.org/10.1007/3-540-45123-4_1}
{doi: {{%
10\hspace{.1pt}\discretionary{.}{%
}{.}\hspace{.4pt}1007\discretionary{/}{%
}{/}3\discretionary{%
}{-}{-}540\discretionary{%
}{-}{-}45123\discretionary{%
}{-}{-}4\_1}}}


\bibitem{BroderCharikarFrieze1998}
A.~Z. Broder, M.~Charikar, A.~M. Frieze, and M.~Mitzenmacher.
\newblock Min-wise independent permutations.
\newblock {\em Journal of Computer and System Sciences}, 60(3):630--659, 2000. \href{https://doi.org/10.1006/jcss.1999.1690}
{doi: {{%
10\hspace{.1pt}\discretionary{.}{%
}{.}\hspace{.4pt}1006\discretionary{/}{%
}{/}jcss\hspace{.1pt}\discretionary{.}{%
}{.}\hspace{.4pt}1999\hspace{.1pt}\discretionary{.}{%
}{.}\hspace{.4pt}1690}}}


\bibitem{CamarriSalvettiBuffoni2005}
S.~Camarri, M.~Buffoni, A.~Iollo, and M.~V. Salvetti.
\newblock Simulation of the three-dimensional flow around a square cylinder between parallel walls at moderate {Reynolds} numbers.
\newblock In {\em XVII Congresso di Meccanica Teorica ed Applicata}, pp. 11--15, 2005. \href{https://doi.org/10.1063/1.869879}
{doi: {{%
10\hspace{.1pt}\discretionary{.}{%
}{.}\hspace{.4pt}1063\discretionary{/}{%
}{/}1\hspace{.1pt}\discretionary{.}{%
}{.}\hspace{.4pt}869879}}}


\bibitem{CardonaMirRossello2013}
G.~Cardona, A.~Mir, F.~Rossell\'{o}, L.~Rotger, and D.~S\'{a}nchez.
\newblock Cophenetic metrics for phylogenetic trees, after {Sokal} and {Rohlf}.
\newblock {\em BMC Bioinformatics}, 14(1):3:1--3:13, 2013. \href{https://doi.org/10.1186/1471-2105-14-3}
{doi: {{%
10\hspace{.1pt}\discretionary{.}{%
}{.}\hspace{.4pt}1186\discretionary{/}{%
}{/}1471\discretionary{%
}{-}{-}2105\discretionary{%
}{-}{-}14\discretionary{%
}{-}{-}3}}}


\bibitem{CarrSnoeyinkAxen2003}
H.~Carr, J.~Snoeyink, and U.~Axen.
\newblock Computing contour trees in all dimensions.
\newblock {\em Computational Geometry: Theory and Applications}, 24(2):75--94, 2003. \href{https://doi.org/10.1016/S0925-7721(02)00093-7}
{doi: {{%
10\hspace{.1pt}\discretionary{.}{%
}{.}\hspace{.4pt}1016\discretionary{/}{%
}{/}S0925\discretionary{%
}{-}{-}7721\discretionary{%
}{(}{(}02\discretionary{)}{%
}{)}00093\discretionary{%
}{-}{-}7}}}


\bibitem{CarrSnoeyinkVanDePanne2010}
H.~Carr, J.~Snoeyink, and M.~Van De~Panne.
\newblock Flexible isosurfaces: Simplifying and displaying scalar topology using the contour tree.
\newblock {\em Computational Geometry}, 43(1):42--58, 2010. \href{https://doi.org/10.1016/j.comgeo.2006.05.009}
{doi: {{%
10\hspace{.1pt}\discretionary{.}{%
}{.}\hspace{.4pt}1016\discretionary{/}{%
}{/}j\hspace{.1pt}\discretionary{.}{%
}{.}\hspace{.4pt}comgeo\hspace{.1pt}\discretionary{.}{%
}{.}\hspace{.4pt}2006\hspace{.1pt}\discretionary{.}{%
}{.}\hspace{.4pt}05\hspace{.1pt}\discretionary{.}{%
}{.}\hspace{.4pt}009}}}


\bibitem{CarrSewellLo2016}
H.~A. Carr, C.~M. Sewell, L.-T. Lo, and J.~P. Ahrens.
\newblock Hybrid data-parallel contour tree computation.
\newblock In {\em Proc. Conferece on Computer Graphics \& Visual Computing}, pp. 73--80, 2016. \href{https://doi.org/10.2312/cgvc.20161299}
{doi: {{%
10\hspace{.1pt}\discretionary{.}{%
}{.}\hspace{.4pt}2312\discretionary{/}{%
}{/}cgvc\hspace{.1pt}\discretionary{.}{%
}{.}\hspace{.4pt}20161299}}}


\bibitem{CarrWeberSewell2019}
H.~A. Carr, G.~H. Weber, C.~M. Sewell, O.~R{\"u}bel, P.~Fasel, and J.~P. Ahrens.
\newblock Scalable contour tree computation by data parallel peak pruning.
\newblock {\em IEEE Transactions on Visualization and Computer Graphics}, 27(4):2437--2454, 2019. \href{https://doi.org/10.1109/TVCG.2019.2948616}
{doi: {{%
10\hspace{.1pt}\discretionary{.}{%
}{.}\hspace{.4pt}1109\discretionary{/}{%
}{/}TVCG\hspace{.1pt}\discretionary{.}{%
}{.}\hspace{.4pt}2019\hspace{.1pt}\discretionary{.}{%
}{.}\hspace{.4pt}2948616}}}


\bibitem{Cglflow2024}
{Computer Graphics Laboratory}.
\newblock \url{https://cgl.ethz.ch/research/visualization/data.php}, 2024.

\bibitem{Charikar2002}
M.~S. Charikar.
\newblock Similarity estimation techniques from rounding algorithms.
\newblock In {\em Proc. 34th {ACM} Symposium on Theory of Computing}, pp. 380--388, 2002. \href{https://doi.org/10.1145/509907.509965}
{doi: {{%
10\hspace{.1pt}\discretionary{.}{%
}{.}\hspace{.4pt}1145\discretionary{/}{%
}{/}509907\hspace{.1pt}\discretionary{.}{%
}{.}\hspace{.4pt}509965}}}


\bibitem{ChenShao2023}
K.~Chen and M.~Shao.
\newblock Locality-sensitive bucketing functions for the edit distance.
\newblock {\em Algorithms for Molecular Biology}, 18(1):7, 2023. \href{https://doi.org/10.1186/s13015-023-00234-2}
{doi: {{%
10\hspace{.1pt}\discretionary{.}{%
}{.}\hspace{.4pt}1186\discretionary{/}{%
}{/}s13015\discretionary{%
}{-}{-}023\discretionary{%
}{-}{-}00234\discretionary{%
}{-}{-}2}}}


\bibitem{ChiLiZhu2014}
L.~Chi, B.~Li, and X.~Zhu.
\newblock Context-preserving hashing for fast text classification.
\newblock In {\em Proc. {SIAM} International Conference on Data Mining}, pp. 100--108, 2014. \href{https://doi.org/10.1137/1.9781611973440.12}
{doi: {{%
10\hspace{.1pt}\discretionary{.}{%
}{.}\hspace{.4pt}1137\discretionary{/}{%
}{/}1\hspace{.1pt}\discretionary{.}{%
}{.}\hspace{.4pt}9781611973440\hspace{.1pt}\discretionary{.}{%
}{.}\hspace{.4pt}12}}}


\bibitem{ChiZhu2017}
L.~Chi and X.~Zhu.
\newblock Hashing techniques: {A} survey and taxonomy.
\newblock {\em ACM Computing Surveys}, 50(1):1--36, 2017. \href{https://doi.org/10.1145/3047307}
{doi: {{%
10\hspace{.1pt}\discretionary{.}{%
}{.}\hspace{.4pt}1145\discretionary{/}{%
}{/}3047307}}}


\bibitem{ChierichettiKumarPanconesi2019}
F.~Chierichetti, R.~Kumar, A.~Panconesi, and E.~Terolli.
\newblock On the distortion of locality sensitive hashing.
\newblock {\em SIAM Journal on Computing}, 48(2):350--372, 2019. \href{https://doi.org/10.1137/17M1127752}
{doi: {{%
10\hspace{.1pt}\discretionary{.}{%
}{.}\hspace{.4pt}1137\discretionary{/}{%
}{/}17M1127752}}}


\bibitem{EdelsbrunnerHarerZomorodian2003}
H.~Edelsbrunner, J.~Harer, and A.~Zomorodian.
\newblock Hierarchical {Morse-Smale} complexes for piecewise linear 2-manifolds.
\newblock {\em Discrete {\&} Computational Geometry}, 30:87--107, 2003. \href{https://doi.org/10.1007/s00454-003-2926-5}
{doi: {{%
10\hspace{.1pt}\discretionary{.}{%
}{.}\hspace{.4pt}1007\discretionary{/}{%
}{/}s00454\discretionary{%
}{-}{-}003\discretionary{%
}{-}{-}2926\discretionary{%
}{-}{-}5}}}


\bibitem{EdelsbrunnerLetscherZomorodian2002}
H.~Edelsbrunner, D.~Letscher, and A.~Zomorodian.
\newblock Topological persistence and simplification.
\newblock {\em Discrete {\&} Computational Geometry}, 28:511--533, 2002. \href{https://doi.org/10.1007/s00454-002-2885-2}
{doi: {{%
10\hspace{.1pt}\discretionary{.}{%
}{.}\hspace{.4pt}1007\discretionary{/}{%
}{/}s00454\discretionary{%
}{-}{-}002\discretionary{%
}{-}{-}2885\discretionary{%
}{-}{-}2}}}


\bibitem{ertl2020}
O.~Ertl.
\newblock {ProbMinHash} -- a class of locality-sensitive hash algorithms for the (probability) {Jaccard} similarity.
\newblock {\em IEEE Transactions on Knowledge and Data Engineering}, 34(7):3491--3506, 2020. \href{https://doi.org/10.1109/TKDE.2020.3021176}
{doi: {{%
10\hspace{.1pt}\discretionary{.}{%
}{.}\hspace{.4pt}1109\discretionary{/}{%
}{/}TKDE\hspace{.1pt}\discretionary{.}{%
}{.}\hspace{.4pt}2020\hspace{.1pt}\discretionary{.}{%
}{.}\hspace{.4pt}3021176}}}


\bibitem{GarofalakisKumar2003}
M.~Garofalakis and A.~Kumar.
\newblock Correlating {XML} data streams using tree-edit distance embeddings.
\newblock In {\em Proc. 22nd ACM SIGMOD-SIGACT-SIGART Symposium on Principles of Database Systems}, pp. 143--154, 2003. \href{https://doi.org/10.1145/773153.773168}
{doi: {{%
10\hspace{.1pt}\discretionary{.}{%
}{.}\hspace{.4pt}1145\discretionary{/}{%
}{/}773153\hspace{.1pt}\discretionary{.}{%
}{.}\hspace{.4pt}773168}}}


\bibitem{GarofalakisKumar2005}
M.~Garofalakis and A.~Kumar.
\newblock {XML} stream processing using tree-edit distance embeddings.
\newblock {\em ACM Transactions on Database Systems}, 30(1):279--332, 2005. \href{https://doi.org/10.1145/1061318.1061326}
{doi: {{%
10\hspace{.1pt}\discretionary{.}{%
}{.}\hspace{.4pt}1145\discretionary{/}{%
}{/}1061318\hspace{.1pt}\discretionary{.}{%
}{.}\hspace{.4pt}1061326}}}


\bibitem{GasparovicMunchOudot2019}
E.~Gasparovic, E.~Munch, S.~Oudot, K.~Turner, B.~Wang, and Y.~Wang.
\newblock Intrinsic interleaving distance for merge trees.
\newblock {\em arXiv eprint ArXiv:1908.00063}, 2019. \href{https://doi.org/10.48550/arXiv.1908.00063}
{doi: {{%
10\hspace{.1pt}\discretionary{.}{%
}{.}\hspace{.4pt}48550\discretionary{/}{%
}{/}arXiv\hspace{.1pt}\discretionary{.}{%
}{.}\hspace{.4pt}1908\hspace{.1pt}\discretionary{.}{%
}{.}\hspace{.4pt}00063}}}


\bibitem{GionisIndykMotwani1999}
A.~Gionis, P.~Indyk, and R.~Motwani.
\newblock Similarity search in high dimensions via hashing.
\newblock In {\em International Conference on Very Large Data Bases (VLDB)}, vol.~99, pp. 518--529, 1999.

\bibitem{GollapudiPanigrahy2008}
S.~Gollapudi and R.~Panigrahy.
\newblock The power of two min-hashes for similarity search among hierarchical data objects.
\newblock In {\em Proc. 27th ACM SIGMOD-SIGACT-SIGART Symposium on Principles of Database Systems}, pp. 211--220, 2008. \href{https://doi.org/10.1145/1376916.1376946}
{doi: {{%
10\hspace{.1pt}\discretionary{.}{%
}{.}\hspace{.4pt}1145\discretionary{/}{%
}{/}1376916\hspace{.1pt}\discretionary{.}{%
}{.}\hspace{.4pt}1376946}}}


\bibitem{GueunetFortinJomier2016}
C.~Gueunet, P.~Fortin, J.~Jomier, and J.~Tierny.
\newblock Contour forests: Fast multi-threaded augmented contour trees.
\newblock In {\em IEEE Symposium on Large Data Analysis and Visualization}, pp. 85--92. IEEE, 2016. \href{https://doi.org/10.1109/LDAV.2016.7874333}
{doi: {{%
10\hspace{.1pt}\discretionary{.}{%
}{.}\hspace{.4pt}1109\discretionary{/}{%
}{/}LDAV\hspace{.1pt}\discretionary{.}{%
}{.}\hspace{.4pt}2016\hspace{.1pt}\discretionary{.}{%
}{.}\hspace{.4pt}7874333}}}


\bibitem{GueunetFortinJomier2017}
C.~Gueunet, P.~Fortin, J.~Jomier, and J.~Tierny.
\newblock Task-based augmented merge trees with {Fibonacci} heaps.
\newblock In {\em IEEE Symposium on Large Data Analysis and Visualization}, pp. 6--15. IEEE, 2017. \href{https://doi.org/10.1109/LDAV.2017.8231846}
{doi: {{%
10\hspace{.1pt}\discretionary{.}{%
}{.}\hspace{.4pt}1109\discretionary{/}{%
}{/}LDAV\hspace{.1pt}\discretionary{.}{%
}{.}\hspace{.4pt}2017\hspace{.1pt}\discretionary{.}{%
}{.}\hspace{.4pt}8231846}}}


\bibitem{GueunetFortinJomier2019}
C.~Gueunet, P.~Fortin, J.~Jomier, and J.~Tierny.
\newblock Task-based augmented contour trees with {Fibonacci} heaps.
\newblock {\em IEEE Transactions on Parallel and Distributed Systems}, 30(8):1889--1905, 2019. \href{https://doi.org/10.1109/TPDS.2019.2898436}
{doi: {{%
10\hspace{.1pt}\discretionary{.}{%
}{.}\hspace{.4pt}1109\discretionary{/}{%
}{/}TPDS\hspace{.1pt}\discretionary{.}{%
}{.}\hspace{.4pt}2019\hspace{.1pt}\discretionary{.}{%
}{.}\hspace{.4pt}2898436}}}


\bibitem{GuntherSalmonTierny2014}
D.~G\"{u}nther, J.~Salmon, and J.~Tierny.
\newblock Mandatory critical points of {2D} uncertain scalar fields.
\newblock {\em Computer Graphics Forum}, 33(3):31--40, 2014. \href{https://doi.org/10.1111/cgf.12359}
{doi: {{%
10\hspace{.1pt}\discretionary{.}{%
}{.}\hspace{.4pt}1111\discretionary{/}{%
}{/}cgf\hspace{.1pt}\discretionary{.}{%
}{.}\hspace{.4pt}12359}}}


\bibitem{GuntherGrossTheisel2017}
T.~G\"{u}nther, M.~Gross, and H.~Theisel.
\newblock Generic objective vortices for flow visualization.
\newblock {\em ACM Transactions on Graphics}, 36(4):141:1--141:11, 2017. \href{https://doi.org/10.1145/3072959.307368}
{doi: {{%
10\hspace{.1pt}\discretionary{.}{%
}{.}\hspace{.4pt}1145\discretionary{/}{%
}{/}3072959\hspace{.1pt}\discretionary{.}{%
}{.}\hspace{.4pt}307368}}}


\bibitem{HeimannLeePan2018}
M.~Heimann, W.~Lee, S.~Pan, K.-Y. Chen, and D.~Koutra.
\newblock {HashAlign:} hash-based alignment of multiple graphs.
\newblock In {\em Advances in Knowledge Discovery and Data Mining}, pp. 726--739. Springer, 2018. \href{https://doi.org/10.1007/978-3-319-93040-4_57}
{doi: {{%
10\hspace{.1pt}\discretionary{.}{%
}{.}\hspace{.4pt}1007\discretionary{/}{%
}{/}978\discretionary{%
}{-}{-}3\discretionary{%
}{-}{-}319\discretionary{%
}{-}{-}93040\discretionary{%
}{-}{-}4\_57}}}


\bibitem{HeineLeitteHlawitschka2016}
C.~Heine, H.~Leitte, M.~Hlawitschka, F.~Iuricich, L.~D. Floriani, G.~Scheuermann, H.~Hagen, and C.~Garth.
\newblock A survey of topology-based methods in visualization.
\newblock {\em Computer Graphics Forum}, 35(3):643--667, 2016. \href{https://doi.org/10.1111/cgf.12933}
{doi: {{%
10\hspace{.1pt}\discretionary{.}{%
}{.}\hspace{.4pt}1111\discretionary{/}{%
}{/}cgf\hspace{.1pt}\discretionary{.}{%
}{.}\hspace{.4pt}12933}}}


\bibitem{IndykMotwani1998}
P.~Indyk and R.~Motwani.
\newblock Approximate nearest neighbors: Towards removing the curse of dimensionality.
\newblock In {\em Proc. 30th Annual {ACM} Symposium on Theory of Computing}, pp. 604--613, 1998. \href{https://doi.org/10.1145/276698.276876}
{doi: {{%
10\hspace{.1pt}\discretionary{.}{%
}{.}\hspace{.4pt}1145\discretionary{/}{%
}{/}276698\hspace{.1pt}\discretionary{.}{%
}{.}\hspace{.4pt}276876}}}


\bibitem{JohanssonMusethCarr2007}
G.~Johansson, K.~Museth, and H.~Carr.
\newblock Flexible and topologically localized segmentation.
\newblock In {\em Eurographics/ IEEE-VGTC Symposium on Visualization}, pp. 179--186, 2007. \href{https://doi.org/10.2312/VisSym/EuroVis07/179-186}
{doi: {{%
10\hspace{.1pt}\discretionary{.}{%
}{.}\hspace{.4pt}2312\discretionary{/}{%
}{/}VisSym\discretionary{/}{%
}{/}EuroVis07\discretionary{/}{%
}{/}179\discretionary{%
}{-}{-}186}}}


\bibitem{LanParsaWang2023}
F.~Lan, S.~Parsa, and B.~Wang.
\newblock Labeled interleaving distance for {Reeb} graphs.
\newblock {\em arXiv preprint arXiv:2306.01186}, 2023. \href{https://doi.org/10.48550/arXiv.2306.01186}
{doi: {{%
10\hspace{.1pt}\discretionary{.}{%
}{.}\hspace{.4pt}48550\discretionary{/}{%
}{/}arXiv\hspace{.1pt}\discretionary{.}{%
}{.}\hspace{.4pt}2306\hspace{.1pt}\discretionary{.}{%
}{.}\hspace{.4pt}01186}}}


\bibitem{LiZhuChi2012}
B.~Li, X.~Zhu, L.~Chi, and C.~Zhang.
\newblock Nested subtree hash kernels for large-scale graph classification over streams.
\newblock In {\em IEEE International Conference on Data Mining}, pp. 399--408. IEEE, 2012. \href{https://doi.org/10.1109/ICDM.2012.101}
{doi: {{%
10\hspace{.1pt}\discretionary{.}{%
}{.}\hspace{.4pt}1109\discretionary{/}{%
}{/}ICDM\hspace{.1pt}\discretionary{.}{%
}{.}\hspace{.4pt}2012\hspace{.1pt}\discretionary{.}{%
}{.}\hspace{.4pt}101}}}


\bibitem{li2022}
H.~Li, W.~Wang, Z.~Liu, Y.~Niu, H.~Wang, S.~Zhao, Y.~Liao, W.~Yang, and X.~Liu.
\newblock A novel locality-sensitive hashing relational graph matching network for semantic textual similarity measurement.
\newblock {\em Expert Systems with Applications}, 207:117832, 2022. \href{https://doi.org/10.1016/j.eswa.2022.117832}
{doi: {{%
10\hspace{.1pt}\discretionary{.}{%
}{.}\hspace{.4pt}1016\discretionary{/}{%
}{/}j\hspace{.1pt}\discretionary{.}{%
}{.}\hspace{.4pt}eswa\hspace{.1pt}\discretionary{.}{%
}{.}\hspace{.4pt}2022\hspace{.1pt}\discretionary{.}{%
}{.}\hspace{.4pt}117832}}}


\bibitem{MarccaisDeBlasioPandey2019}
G.~Mar{\c{c}}ais, D.~DeBlasio, P.~Pandey, and C.~Kingsford.
\newblock Locality-sensitive hashing for the edit distance.
\newblock {\em Bioinformatics}, 35(14):i127--i135, 2019. \href{https://doi.org/10.1093/bioinformatics/btz354}
{doi: {{%
10\hspace{.1pt}\discretionary{.}{%
}{.}\hspace{.4pt}1093\discretionary{/}{%
}{/}bioinformatics\discretionary{/}{%
}{/}btz354}}}


\bibitem{Mccauley2021}
S.~McCauley.
\newblock Approximate similarity search under edit distance using locality-sensitive hashing.
\newblock In {\em 24th International Conference on Database Theory}, pp. 21:1--21:22, 2021. \href{https://doi.org/10.4230/LIPIcs.ICDT.2021.21}
{doi: {{%
10\hspace{.1pt}\discretionary{.}{%
}{.}\hspace{.4pt}4230\discretionary{/}{%
}{/}LIPIcs\hspace{.1pt}\discretionary{.}{%
}{.}\hspace{.4pt}ICDT\hspace{.1pt}\discretionary{.}{%
}{.}\hspace{.4pt}2021\hspace{.1pt}\discretionary{.}{%
}{.}\hspace{.4pt}21}}}


\bibitem{MizutaMatsuda2005}
S.~Mizuta and T.~Matsuda.
\newblock Description of digital images by region-based contour trees.
\newblock In {\em International Conference Image Analysis and Recognition}, pp. 549--558. Springer, 2005. \href{https://doi.org/10.1007/11559573_68}
{doi: {{%
10\hspace{.1pt}\discretionary{.}{%
}{.}\hspace{.4pt}1007\discretionary{/}{%
}{/}11559573\_68}}}


\bibitem{MorozovBeketayevWeber2013}
D.~Morozov, K.~Beketayev, and G.~Weber.
\newblock Interleaving distance between merge trees.
\newblock {\em Topology-Based Methods in Visualization}, 2013.

\bibitem{NarayananThomasNatarajan2015}
V.~Narayanan, D.~M. Thomas, and V.~Natarajan.
\newblock Distance between extremum graphs.
\newblock In {\em IEEE Pacific Visualization Symposium}, pp. 263--270, 2015. \href{https://doi.org/10.1109/PACIFICVIS.2015.7156386}
{doi: {{%
10\hspace{.1pt}\discretionary{.}{%
}{.}\hspace{.4pt}1109\discretionary{/}{%
}{/}PACIFICVIS\hspace{.1pt}\discretionary{.}{%
}{.}\hspace{.4pt}2015\hspace{.1pt}\discretionary{.}{%
}{.}\hspace{.4pt}7156386}}}


\bibitem{OesterlingHeineJanicke2011}
P.~Oesterling, C.~Heine, H.~Janicke, G.~Scheuermann, and G.~Heyer.
\newblock Visualization of high-dimensional point clouds using their density distribution's topology.
\newblock {\em IEEE Transactions on Visualization and Computer Graphics}, 17(11):1547--1559, 2011. \href{https://doi.org/10.1109/TVCG.2011.27}
{doi: {{%
10\hspace{.1pt}\discretionary{.}{%
}{.}\hspace{.4pt}1109\discretionary{/}{%
}{/}TVCG\hspace{.1pt}\discretionary{.}{%
}{.}\hspace{.4pt}2011\hspace{.1pt}\discretionary{.}{%
}{.}\hspace{.4pt}27}}}


\bibitem{PocoDoraiswamyTalbert2015}
J.~Poco, H.~Doraiswamy, M.~Talbert, J.~Morisette, and C.~T. Silva.
\newblock Using maximum topology matching to explore differences in species distribution models.
\newblock {\em IEEE Scientific Visualization Conference}, pp. 9--16, 2015. \href{https://doi.org/10.1109/SciVis.2015.7429486}
{doi: {{%
10\hspace{.1pt}\discretionary{.}{%
}{.}\hspace{.4pt}1109\discretionary{/}{%
}{/}SciVis\hspace{.1pt}\discretionary{.}{%
}{.}\hspace{.4pt}2015\hspace{.1pt}\discretionary{.}{%
}{.}\hspace{.4pt}7429486}}}


\bibitem{PontVidalDelon2022}
M.~Pont, J.~Vidal, J.~Delon, and J.~Tierny.
\newblock Wasserstein distances, geodesics and barycenters of merge trees.
\newblock {\em IEEE Transactions on Visualization and Computer Graphics}, 28(1):291--301, 2022. \href{https://doi.org/10.1109/TVCG.2021.3114839}
{doi: {{%
10\hspace{.1pt}\discretionary{.}{%
}{.}\hspace{.4pt}1109\discretionary{/}{%
}{/}TVCG\hspace{.1pt}\discretionary{.}{%
}{.}\hspace{.4pt}2021\hspace{.1pt}\discretionary{.}{%
}{.}\hspace{.4pt}3114839}}}


\bibitem{gerrisflowsolver}
S.~Popinet.
\newblock Free computational fluid dynamics.
\newblock {\em ClusterWorld}, 2(6), 2004.

\bibitem{QinFasyWenk2021}
Y.~Qin, B.~T. Fasy, C.~Wenk, and B.~Summa.
\newblock A domain-oblivious approach for learning concise representations of filtered topological spaces for clustering.
\newblock {\em IEEE Transactions on Visualization and Computer Graphics}, 28(1):302--312, 2021. \href{https://doi.org/10.1109/TVCG.2021.3114872}
{doi: {{%
10\hspace{.1pt}\discretionary{.}{%
}{.}\hspace{.4pt}1109\discretionary{/}{%
}{/}TVCG\hspace{.1pt}\discretionary{.}{%
}{.}\hspace{.4pt}2021\hspace{.1pt}\discretionary{.}{%
}{.}\hspace{.4pt}3114872}}}


\bibitem{RosenSethMills2021}
P.~Rosen, A.~Seth, E.~Mills, A.~Ginsburg, J.~Kamenetzky, J.~Kern, C.~R. Johnson, and B.~Wang.
\newblock Using contour trees in the analysis and visualization of radio astronomy data cubes.
\newblock In {\em Topological Methods in Data Analysis and Visualization VI}, pp. 87--108, 2021. \href{https://doi.org/10.1007/978-3-030-83500-2_6}
{doi: {{%
10\hspace{.1pt}\discretionary{.}{%
}{.}\hspace{.4pt}1007\discretionary{/}{%
}{/}978\discretionary{%
}{-}{-}3\discretionary{%
}{-}{-}030\discretionary{%
}{-}{-}83500\discretionary{%
}{-}{-}2\_6}}}


\bibitem{SaikiaSeidelWeinkauf2014}
H.~Saikia, H.-P. Seidel, and T.~Weinkauf.
\newblock Extended branch decomposition graphs: Structural comparison of scalar data.
\newblock {\em Computer Graphics Forum}, 33(3):41--50, 2014. \href{https://doi.org/10.1111/cgf.12360}
{doi: {{%
10\hspace{.1pt}\discretionary{.}{%
}{.}\hspace{.4pt}1111\discretionary{/}{%
}{/}cgf\hspace{.1pt}\discretionary{.}{%
}{.}\hspace{.4pt}12360}}}


\bibitem{SaikiaSeidelWeinkauf2015}
H.~Saikia, H.-P. Seidel, and T.~Weinkauf.
\newblock Fast similarity search in scalar fields using merging histograms.
\newblock In {\em Topological Methods in Data Analysis and Visualization IV}, pp. 121--134, 2017. \href{https://doi.org/10.1007/978-3-319-44684-4_7}
{doi: {{%
10\hspace{.1pt}\discretionary{.}{%
}{.}\hspace{.4pt}1007\discretionary{/}{%
}{/}978\discretionary{%
}{-}{-}3\discretionary{%
}{-}{-}319\discretionary{%
}{-}{-}44684\discretionary{%
}{-}{-}4\_7}}}


\bibitem{SaikiaWeinkauf2017}
H.~Saikia and T.~Weinkauf.
\newblock Global feature tracking and similarity estimation in time-dependent scalar fields.
\newblock {\em Computer Graphics Forum}, 36(3):1--11, 2017. \href{https://doi.org/10.1111/cgf.13163}
{doi: {{%
10\hspace{.1pt}\discretionary{.}{%
}{.}\hspace{.4pt}1111\discretionary{/}{%
}{/}cgf\hspace{.1pt}\discretionary{.}{%
}{.}\hspace{.4pt}13163}}}


\bibitem{Viscousfingers2016}
Scientific visualization contest.
\newblock \url{http://www.uni-kl.de/sciviscontest/}, 2016.

\bibitem{TOSCA2024}
{TOSCA}.
\newblock \url{http://tosca.cs.technion.ac.il/}, 2024.

\bibitem{ShervashidzeSchweitzerVanLeeuwen2011}
N.~Shervashidze, P.~Schweitzer, E.~J. Van~Leeuwen, K.~Mehlhorn, and K.~M. Borgwardt.
\newblock {Weisfeiler-Lehman} graph kernels.
\newblock {\em Journal of Machine Learning Research}, 12(9), 2011.

\bibitem{ShinIshikawa2018}
K.~Shin and T.~Ishikawa.
\newblock Linear-time algorithms for the subpath kernel.
\newblock In {\em Proc. 29th Annual Symposium on Combinatorial Pattern Matching}, pp. 22:1--22:13. Schloss Dagstuhl-Leibniz-Zentrum fuer Informatik, 2018. \href{https://doi.org/10.4230/LIPIcs.CPM.2018.22}
{doi: {{%
10\hspace{.1pt}\discretionary{.}{%
}{.}\hspace{.4pt}4230\discretionary{/}{%
}{/}LIPIcs\hspace{.1pt}\discretionary{.}{%
}{.}\hspace{.4pt}CPM\hspace{.1pt}\discretionary{.}{%
}{.}\hspace{.4pt}2018\hspace{.1pt}\discretionary{.}{%
}{.}\hspace{.4pt}22}}}


\bibitem{SolerPlainchaultConche2018}
M.~Soler, M.~Plainchault, B.~Conche, and J.~Tierny.
\newblock Lifted {Wasserstein} matcher for fast and robust topology tracking.
\newblock In {\em IEEE Symposium on Large Data Analysis and Visualization}, pp. 23--33, 2018. \href{https://doi.org/10.1109/LDAV.2018.8739196}
{doi: {{%
10\hspace{.1pt}\discretionary{.}{%
}{.}\hspace{.4pt}1109\discretionary{/}{%
}{/}LDAV\hspace{.1pt}\discretionary{.}{%
}{.}\hspace{.4pt}2018\hspace{.1pt}\discretionary{.}{%
}{.}\hspace{.4pt}8739196}}}


\bibitem{SridharamurthyMasoodKamakshidasan2020}
R.~Sridharamurthy, T.~B. Masood, A.~Kamakshidasan, and V.~Natarajan.
\newblock Edit distance between merge trees.
\newblock {\em {IEEE} Transactions on Visualization and Computer Graphics}, 26(3):1518--1531, 2020. \href{https://doi.org/10.1109/TVCG.2018.2873612}
{doi: {{%
10\hspace{.1pt}\discretionary{.}{%
}{.}\hspace{.4pt}1109\discretionary{/}{%
}{/}TVCG\hspace{.1pt}\discretionary{.}{%
}{.}\hspace{.4pt}2018\hspace{.1pt}\discretionary{.}{%
}{.}\hspace{.4pt}2873612}}}


\bibitem{SridharamurthyNatarajan2023}
R.~Sridharamurthy and V.~Natarajan.
\newblock Comparative analysis of merge trees using local tree edit distance.
\newblock {\em IEEE Transactions on Visualization and Computer Graphics}, 29(2):1518--1530, 2023. \href{https://doi.org/10.1109/TVCG.2021.3122176}
{doi: {{%
10\hspace{.1pt}\discretionary{.}{%
}{.}\hspace{.4pt}1109\discretionary{/}{%
}{/}TVCG\hspace{.1pt}\discretionary{.}{%
}{.}\hspace{.4pt}2021\hspace{.1pt}\discretionary{.}{%
}{.}\hspace{.4pt}3122176}}}


\bibitem{TatikondaParthasarathy2010}
S.~Tatikonda and S.~Parthasarathy.
\newblock Hashing tree-structured data: Methods and applications.
\newblock In {\em IEEE International Conference on Data Engineering}, pp. 429--440. IEEE, 2010. \href{https://doi.org/10.1109/ICDE.2010.5447882}
{doi: {{%
10\hspace{.1pt}\discretionary{.}{%
}{.}\hspace{.4pt}1109\discretionary{/}{%
}{/}ICDE\hspace{.1pt}\discretionary{.}{%
}{.}\hspace{.4pt}2010\hspace{.1pt}\discretionary{.}{%
}{.}\hspace{.4pt}5447882}}}


\bibitem{ThomasNatarajan2014}
D.~M. Thomas and V.~Natarajan.
\newblock Multiscale symmetry detection in scalar fields by clustering contours.
\newblock {\em IEEE Transactions on Visualization and Computer Graphics}, 20(12):2427--2436, 2014. \href{https://doi.org/10.1109/TVCG.2014.2346332}
{doi: {{%
10\hspace{.1pt}\discretionary{.}{%
}{.}\hspace{.4pt}1109\discretionary{/}{%
}{/}TVCG\hspace{.1pt}\discretionary{.}{%
}{.}\hspace{.4pt}2014\hspace{.1pt}\discretionary{.}{%
}{.}\hspace{.4pt}2346332}}}


\bibitem{TiernyFavelierLevine2018}
J.~Tierny, G.~Favelier, J.~A. Levine, C.~Gueunet, and M.~Michaux.
\newblock The {Topology ToolKit}.
\newblock {\em {IEEE} Transactions on Visualization and Computer Graphics}, 24(1):832--842, 2018. \href{https://doi.org/10.1109/TVCG.2017.2743938}
{doi: {{%
10\hspace{.1pt}\discretionary{.}{%
}{.}\hspace{.4pt}1109\discretionary{/}{%
}{/}TVCG\hspace{.1pt}\discretionary{.}{%
}{.}\hspace{.4pt}2017\hspace{.1pt}\discretionary{.}{%
}{.}\hspace{.4pt}2743938}}}


\bibitem{WeberDillardCarr2007}
G.~H. Weber, S.~E. Dillard, H.~Carr, V.~Pascucci, and B.~Hamann.
\newblock Topology-controlled volume rendering.
\newblock {\em IEEE Transactions on Visualization and Computer Graphics}, 13(2):330--341, 2007. \href{https://doi.org/10.1109/TVCG.2007.47}
{doi: {{%
10\hspace{.1pt}\discretionary{.}{%
}{.}\hspace{.4pt}1109\discretionary{/}{%
}{/}TVCG\hspace{.1pt}\discretionary{.}{%
}{.}\hspace{.4pt}2007\hspace{.1pt}\discretionary{.}{%
}{.}\hspace{.4pt}47}}}


\bibitem{WernerGarth2021}
K.~Werner and C.~Garth.
\newblock Unordered task-parallel augmented merge tree construction.
\newblock {\em IEEE Transactions on Visualization and Computer Graphics}, 27(8):3585--3596, 2021. \href{https://doi.org/10.1109/TVCG.2021.3076875}
{doi: {{%
10\hspace{.1pt}\discretionary{.}{%
}{.}\hspace{.4pt}1109\discretionary{/}{%
}{/}TVCG\hspace{.1pt}\discretionary{.}{%
}{.}\hspace{.4pt}2021\hspace{.1pt}\discretionary{.}{%
}{.}\hspace{.4pt}3076875}}}


\bibitem{WetzelsAndersGarth2023}
F.~Wetzels, M.~Anders, and C.~Garth.
\newblock Taming horizontal instability in merge trees: On the computation of a comprehensive deformation-based edit distance.
\newblock In {\em Topological Data Analysis and Visualization}, pp. 82--92, 2023. \href{https://doi.org/10.1109/TopoInVis60193.2023.00015}
{doi: {{%
10\hspace{.1pt}\discretionary{.}{%
}{.}\hspace{.4pt}1109\discretionary{/}{%
}{/}TopoInVis60193\hspace{.1pt}\discretionary{.}{%
}{.}\hspace{.4pt}2023\hspace{.1pt}\discretionary{.}{%
}{.}\hspace{.4pt}00015}}}


\bibitem{WetzelsGarth2022}
F.~Wetzels and C.~Garth.
\newblock A deformation-based edit distance for merge trees.
\newblock In {\em Topological Data Analysis and Visualization}, pp. 29--38, 2022. \href{https://doi.org/10.1109/TopoInVis57755.2022.00010}
{doi: {{%
10\hspace{.1pt}\discretionary{.}{%
}{.}\hspace{.4pt}1109\discretionary{/}{%
}{/}TopoInVis57755\hspace{.1pt}\discretionary{.}{%
}{.}\hspace{.4pt}2022\hspace{.1pt}\discretionary{.}{%
}{.}\hspace{.4pt}00010}}}


\bibitem{WetzelsLeitteGarth2022}
F.~Wetzels, H.~Leitte, and C.~Garth.
\newblock Branch decomposition-independent edit distances for merge trees.
\newblock {\em {Computer Graphics Forum}}, 41(3):367--378, 2022. \href{https://doi.org/10.1111/cgf.14547}
{doi: {{%
10\hspace{.1pt}\discretionary{.}{%
}{.}\hspace{.4pt}1111\discretionary{/}{%
}{/}cgf\hspace{.1pt}\discretionary{.}{%
}{.}\hspace{.4pt}14547}}}


\bibitem{WidanagamaachchiJacquesWang2017}
W.~Widanagamaachchi, A.~Jacques, B.~Wang, E.~Crosman, P.-T. Bremer, V.~Pascucci, and J.~Horel.
\newblock Exploring the evolution of pressure-perturbations to understand atmospheric phenomena.
\newblock In {\em IEEE Pacific Visualization Symposium}, pp. 101--110, 2017. \href{https://doi.org/10.1109/PACIFICVIS.2017.8031584}
{doi: {{%
10\hspace{.1pt}\discretionary{.}{%
}{.}\hspace{.4pt}1109\discretionary{/}{%
}{/}PACIFICVIS\hspace{.1pt}\discretionary{.}{%
}{.}\hspace{.4pt}2017\hspace{.1pt}\discretionary{.}{%
}{.}\hspace{.4pt}8031584}}}


\bibitem{WoodHoppeDesbrun2004}
Z.~Wood, H.~Hoppe, M.~Desbrun, and P.~Schr{\"o}der.
\newblock Removing excess topology from isosurfaces.
\newblock {\em ACM Transactions on Graphics}, 23(2):190--208, 2004. \href{https://doi.org/10.1145/990002.990007}
{doi: {{%
10\hspace{.1pt}\discretionary{.}{%
}{.}\hspace{.4pt}1145\discretionary{/}{%
}{/}990002\hspace{.1pt}\discretionary{.}{%
}{.}\hspace{.4pt}990007}}}


\bibitem{WuZhang2013}
K.~Wu and S.~Zhang.
\newblock A contour tree based visualization for exploring data with uncertainty.
\newblock {\em International Journal for Uncertainty Quantification}, 3(3), 2013. \href{https://doi.org/10.1615/Int.J.UncertaintyQuantification.2012003956}
{doi: {{%
10\hspace{.1pt}\discretionary{.}{%
}{.}\hspace{.4pt}1615\discretionary{/}{%
}{/}Int\hspace{.1pt}\discretionary{.}{%
}{.}\hspace{.4pt}J\hspace{.1pt}\discretionary{.}{%
}{.}\hspace{.4pt}UncertaintyQuantification\hspace{.1pt}\discretionary{.}{%
}{.}\hspace{.4pt}2012003956}}}


\bibitem{WuLi2022}
W.~Wu and B.~Li.
\newblock Locality sensitive hashing for structured data: A survey.
\newblock {\em arXiv preprint arXiv:2204.11209}, 2022. \href{https://doi.org/10.48550/arXiv.2204.11209}
{doi: {{%
10\hspace{.1pt}\discretionary{.}{%
}{.}\hspace{.4pt}48550\discretionary{/}{%
}{/}arXiv\hspace{.1pt}\discretionary{.}{%
}{.}\hspace{.4pt}2204\hspace{.1pt}\discretionary{.}{%
}{.}\hspace{.4pt}11209}}}


\bibitem{WuLiChen2020}
W.~Wu, B.~Li, L.~Chen, J.~Gao, and C.~Zhang.
\newblock A review for weighted {MinHash} algorithms.
\newblock {\em IEEE Transactions on Knowledge and Data Engineering}, 34(6):2553--2573, 2020. \href{https://doi.org/10.1109/TKDE.2020.3021067}
{doi: {{%
10\hspace{.1pt}\discretionary{.}{%
}{.}\hspace{.4pt}1109\discretionary{/}{%
}{/}TKDE\hspace{.1pt}\discretionary{.}{%
}{.}\hspace{.4pt}2020\hspace{.1pt}\discretionary{.}{%
}{.}\hspace{.4pt}3021067}}}


\bibitem{WuLiChen2017}
W.~Wu, B.~Li, L.~Chen, X.~Zhu, and C.~Zhang.
\newblock {$K$}-ary tree hashing for fast graph classification.
\newblock {\em IEEE Transactions on Knowledge and Data Engineering}, 30(5):936--949, 2017. \href{https://doi.org/10.1109/TKDE.2017.2782278}
{doi: {{%
10\hspace{.1pt}\discretionary{.}{%
}{.}\hspace{.4pt}1109\discretionary{/}{%
}{/}TKDE\hspace{.1pt}\discretionary{.}{%
}{.}\hspace{.4pt}2017\hspace{.1pt}\discretionary{.}{%
}{.}\hspace{.4pt}2782278}}}


\bibitem{XuNiuJi2022}
Z.~Xu, L.~Niu, J.~Ji, and Q.~Li.
\newblock Structure-preserving hashing for tree-structured data.
\newblock {\em Signal, Image and Video Processing}, 16(8):2045--2053, 2022. \href{https://doi.org/10.1007/s11760-022-02166-7}
{doi: {{%
10\hspace{.1pt}\discretionary{.}{%
}{.}\hspace{.4pt}1007\discretionary{/}{%
}{/}s11760\discretionary{%
}{-}{-}022\discretionary{%
}{-}{-}02166\discretionary{%
}{-}{-}7}}}


\bibitem{YanGuoPeterka2024}
L.~Yan, H.~Guo, T.~Peterka, B.~Wang, and J.~Wang.
\newblock {TROPHY}: A topologically robust physics-informed tracking framework for tropical cyclones.
\newblock {\em IEEE Transactions on Visualization and Computer Graphics}, 30:1302--1312, 2024. \href{https://doi.org/10.1109/TVCG.2023.3326905}
{doi: {{%
10\hspace{.1pt}\discretionary{.}{%
}{.}\hspace{.4pt}1109\discretionary{/}{%
}{/}TVCG\hspace{.1pt}\discretionary{.}{%
}{.}\hspace{.4pt}2023\hspace{.1pt}\discretionary{.}{%
}{.}\hspace{.4pt}3326905}}}


\bibitem{YanMasoodRasheed2023}
L.~Yan, T.~B. Masood, F.~Rasheed, I.~Hotz, and B.~Wang.
\newblock Geometry-aware merge tree comparisons for time-varying data with interleaving distances.
\newblock {\em IEEE Transactions on Visualization and Computer Graphics}, 29(8):3489--3506, 2023. \href{https://doi.org/10.1109/TVCG.2022.3163349}
{doi: {{%
10\hspace{.1pt}\discretionary{.}{%
}{.}\hspace{.4pt}1109\discretionary{/}{%
}{/}TVCG\hspace{.1pt}\discretionary{.}{%
}{.}\hspace{.4pt}2022\hspace{.1pt}\discretionary{.}{%
}{.}\hspace{.4pt}3163349}}}


\bibitem{YanMasoodSridharamurthy2021}
L.~Yan, T.~B. Masood, R.~Sridharamurthy, F.~Rasheed, V.~Natarajan, I.~Hotz, and B.~Wang.
\newblock Scalar field comparison with topological descriptors: Properties and applications for scientific visualization.
\newblock {\em Computer Graphics Forum}, 40(3):599--633, 2021. \href{https://doi.org/10.1111/cgf.14331}
{doi: {{%
10\hspace{.1pt}\discretionary{.}{%
}{.}\hspace{.4pt}1111\discretionary{/}{%
}{/}cgf\hspace{.1pt}\discretionary{.}{%
}{.}\hspace{.4pt}14331}}}


\bibitem{YanWangMunch2020}
L.~Yan, Y.~Wang, E.~Munch, E.~Gasparovic, and B.~Wang.
\newblock A structural average of labeled merge trees for uncertainty visualization.
\newblock {\em IEEE Transactions on Visualization and Computer Graphics}, 26(1):832--842, 2020. \href{https://doi.org/10.1109/TVCG.2019.2934242}
{doi: {{%
10\hspace{.1pt}\discretionary{.}{%
}{.}\hspace{.4pt}1109\discretionary{/}{%
}{/}TVCG\hspace{.1pt}\discretionary{.}{%
}{.}\hspace{.4pt}2019\hspace{.1pt}\discretionary{.}{%
}{.}\hspace{.4pt}2934242}}}


\bibitem{ZhangZhang2017}
H.~Zhang and Q.~Zhang.
\newblock {Embedjoin:} efficient edit similarity joins via embeddings.
\newblock In {\em Proc. 23rd {ACM SIGKDD} International Conference on Knowledge Discovery \& Data Mining}, pp. 585--594, 2017. \href{https://doi.org/10.1145/3097983.3098003}
{doi: {{%
10\hspace{.1pt}\discretionary{.}{%
}{.}\hspace{.4pt}1145\discretionary{/}{%
}{/}3097983\hspace{.1pt}\discretionary{.}{%
}{.}\hspace{.4pt}3098003}}}


\bibitem{ZhangZhang2019}
H.~Zhang and Q.~Zhang.
\newblock {MinJoin:} efficient edit similarity joins via local hash minima.
\newblock In {\em Proc. 25th {ACM SIGKDD} International Conference on Knowledge Discovery \& Data Mining}, pp. 1093--1103, 2019. \href{https://doi.org/10.1145/3292500.3330853}
{doi: {{%
10\hspace{.1pt}\discretionary{.}{%
}{.}\hspace{.4pt}1145\discretionary{/}{%
}{/}3292500\hspace{.1pt}\discretionary{.}{%
}{.}\hspace{.4pt}3330853}}}


\bibitem{ZhangZhang2020}
H.~Zhang and Q.~Zhang.
\newblock {MinSearch:} an efficient algorithm for similarity search under edit distance.
\newblock In {\em Proc. 26th {ACM SIGKDD} International Conference on Knowledge Discovery \& Data Mining}, pp. 566--576, 2020. \href{https://doi.org/10.1145/3394486.3403099}
{doi: {{%
10\hspace{.1pt}\discretionary{.}{%
}{.}\hspace{.4pt}1145\discretionary{/}{%
}{/}3394486\hspace{.1pt}\discretionary{.}{%
}{.}\hspace{.4pt}3403099}}}


\end{thebibliography}
\clearpage
\newpage
\appendix 
\section{Pseudocode}
\label{sec:pesudocode}

We provide pseudocode for the relevant algorithms described in the paper. 
\cref{algorithm:RMH} presents the pseudocode for the recursive MinHash. 
\cref{algorithm:dR} presents the pseudocode for the modified recursive MinHash for merge trees. 
\cref{algorithm:signature-visit} presents the pseudocode for storing subpaths using modified DFS. 
\cref{algorithm:signature} presents the pseudocode for generating subpath signatures for merge trees.

\begin{algorithm}
\caption{RMH}
\label{algorithm:RMH}
\LinesNumbered
\SetAlgoLined
\DontPrintSemicolon
\KwData{Nested set $W^{(r)}$ at $r$th level, $k$ hash functions at $r$th level, $\{\sigma^{(r)}_i\}^{R,k}_{r=1, i=1}$}
\KwResult{$\textbf{h}^{(r)}$, fingerprint of $W^{(r)}$}
\Begin{
\tcp{Level 1}
\If{r = 1}
{$\textbf{h}^{(r)} \leftarrow \mathrm{MinHash}(W^{(r)},\{(\sigma^{(r)}_i)\}^{k}_{i=1})$}
\Else{
\For{$W^{(r-1)}_{(*)} \in W^{(r)}$}
{
\tcp{Higher levels}
$\textbf{h}^{(r)}_{*} \leftarrow \mathrm{RMH}(W^{(r-1)}_{(*)},\{(\sigma^{(r)}_i)\}^{R,k}_{r=1, i=1})$
}
$M \leftarrow |W^{(r)}|$\;
$L = dim(\textbf{h}^{(r-1)}_{*})$\;
\tcp{Reorganize}
\For{$l\leftarrow 1$ \KwTo $L$}{
$W^{(r)}_{(l)} = \{h^{r-1}_{m, (l)}\}^{M}_{m=1} $\;
$\textbf{h}^{(r)}_{(l)} \leftarrow \mathrm{MinHash}(W^{(r)}_{(l)},\{(\sigma^{(r)}_i)\}^{k}_{i=1})$
}
}
\tcp{Concatenate}
$\textbf{h}^{(r)} = [\textbf{h}^{(r)}_{(1)}; \textbf{h}^{(r)}_{(2)}; \ldots ;\textbf{h}^{(r)}_{(L)}]$
}

\end{algorithm}

\begin{algorithm}[!ht]
\caption{RMH Signatures for Merge Trees}
\label{algorithm:dR}
\LinesNumbered
\SetAlgoLined
\DontPrintSemicolon
\KwData{Set of nodes at $r$th level i.e ${T}^{(r)}$ , $K$ hash functions at $r$th level,  i.e.,~$\{\sigma^{(r)}_k\}^{R,K}_{r=1, k=1}$}
\KwResult{$\textbf{h}^{(r)}$, fingerprint of ${T}^{(r)}$}

\Begin{
\For{${T}^{(r)}_{(*)} \in {T}^{(r)}$}
{
\tcp{Extrema}
\If{$deg({T}^{(r)}_{(*)}) = 1$}{
$\textbf{h}^{(r)}_{*} \leftarrow \mathrm{MinHash}({T}^{(r)}_{(*)},\{(\sigma^{(r)}_k)\}^{K}_{k=1})$
}
\Else{
\tcp{Saddles recursive call}
$\textbf{h}^{(r)}_{*} \leftarrow \mathrm{RMH}({T}^{(r)}_{(*)},\{(\sigma^{(r)}_k)\}^{R,K}_{r=1, k=1})$
}
}
$M \leftarrow |{T}^{(r)}|$\;
$L = dim(\textbf{h}^{(r-1)}_{*})$\;
\tcp{Reorganize}
\For{$l\leftarrow 1$ \KwTo $L$}{
${T}^{(r)}_{(l)} = \{h^{r-1}_{m, (l)}\}^{M}_{m=1} $\;
$\textbf{h}^{(r)}_{(l)} \leftarrow \mathrm{MinHash}({T}^{(r)}_{(l)},\{(\sigma^{(r)}_k)\}^{K}_{k=1})$
}

\tcp{Concatenate}
$\textbf{h}^{(r)} = [\textbf{h}^{(r)}_{(1)}; \textbf{h}^{(r)}_{(2)}; \ldots ;\textbf{h}^{(r)}_{(L)}]$
}

\end{algorithm}

\begin{algorithm}
\caption{SS-VISIT}
\label{algorithm:signature-visit}
\LinesNumbered
\SetAlgoLined
\DontPrintSemicolon
\KwData{$u \in T$, multiset $SP$ to store subpaths}
\KwResult{Builds $SP$}
\Begin{
$\mathsf{PUSH(S,u)}$\;
$color[u] \leftarrow gray$\;
\For{$v \in children(u)$}{
\If{$color[v] = white$}{
$\mathsf{SS}-\mathsf{VISIT(v)}$
}
}
$color[u] \leftarrow black$\;
$sp \leftarrow \mathsf{POP}-\mathsf{PUSH(t,t-1)}$\;
$sp \leftarrow \mathsf{REVERSE(sp)}$\;
$SP \leftarrow SP + sp$
}
\end{algorithm}

\begin{algorithm}
\caption{SS Signatures for Merge Trees}
\label{algorithm:signature}
\LinesNumbered
\SetAlgoLined
\DontPrintSemicolon
\KwData{Tree ${T}$ , subpath length $t$}
\KwResult{Multiset $SP$ containing all subpaths of length $t$.}
\Begin{
$D \leftarrow \{d_1,d_2, \ldots , d_{t-1}\}$
$T' \leftarrow \mathsf{ConcatenateDummyNodes(D, T)}$\;
$S \leftarrow \varnothing$\;
\For{$u \in V(T')$}
{
$color[u] \leftarrow white$
}
$\mathsf{SS-VISIT(root(T'))}$\;
\Return $SP$
}
\end{algorithm}
\section{Hash Functions}
\label{sec:hash-function}

We provide a brief description of how hashes used in \cref{fig:text-rmh} are generated using random permutations.
The illustrative example in \cref{fig:text-rmh} shows that RMH can be constructed as follows. 

Assume a universal set $U = \{a,b,c, \ldots , j\}$ which can be indexed by a set $I = \{1,2,3, \ldots, 10\}$. 
Now the sets $S_1 = \{{a,b,c}\}, S_2=\{{b,e}\}, S_3=\{{d,e,a}\}$ can be represented using a $0$-$1$ encoding as shown in~\cref{table:zoencode1}.

\begin{table}[ht]
\centering
\begin{tabular}{|l|l|lll|}
\hline
U & I & \textbf{$S_1$} & \textbf{$S_2$} & \textbf{$S_3$} \\ \hline
$a$  & 1  & 1     & 0     & 1     \\
$b$  & 2  & 1     & 1     & 0     \\
$c$  & 3  & 1     & 0     & 0     \\
$d$  & 4  & 0     & 0     & 1     \\
$e$  & 5  & 0     & 1     & 1     \\
$f$  & 6  & 0     & 0     & 0     \\
$g$  & 7  & 0     & 0     & 0     \\
$h$  & 8  & 0     & 0     & 0     \\
$i$  & 9  & 0     & 0     & 0     \\
$j$  & 10 & 0     & 0     & 0     \\
\hline
\end{tabular}
\caption{Sets $S_1,S_2,S_3$ represented using a $0$-$1$ encoding along with the universal set $U$ and the index set $I$.}
\label{table:zoencode1}
\vspace{-4mm}
\end{table}

We generate four random permutations $\pi_1, \pi_2, \pi_3, \pi_4$ on the index set $I$ as follows, 
\[
\pi_1\{I\} = \{3,2,4,1,5,6,7,8,9,10\}, 
\]
\[
\pi_2\{I\} = \{4,5,3,6,8,9,10,2,1,7\}, 
\]
\[
\pi_3\{I\} = \{1,3,5,2,4,7,9,6,8,10\}, 
\]
\[
\pi_4\{I\} = \{1,4,7,6,5,3,8,2,9,10\}. 
\]
These permutations provide us four hash functions $h_1, h_2, h_3$, and $h_4$. We show results for $h_1$, the rest of the results are similar.

\begin{table}[ht]
\centering
\begin{tabular}{|l|l|lll|}
\hline
U & I & \textbf{$S_1$} & \textbf{$S_2$} & \textbf{$S_3$} \\ \hline
$c$  & 3  & 1     & 0     & 0     \\
$b$  & 2  & 1     & 1     & 0     \\
$d$  & 4  & 0     & 0     & 1     \\
$a$  & 1  & 1     & 0     & 1     \\
$e$  & 5  & 0     & 1     & 1     \\
$f$  & 6  & 0     & 0     & 0     \\
$g$  & 7  & 0     & 0     & 0     \\
$h$  & 8  & 0     & 0     & 0     \\
$i$  & 9  & 0     & 0     & 0     \\
$j$  & 10 & 0     & 0     & 0     \\
\hline
\end{tabular}
\caption{permutation $\pi_1$ resulting in hash function $h_1$.}
\label{table:hash1}
\end{table}

MinHash stores the index of the first occurrence of $1$ in the particular permutation for each of the sets.~\cref{table:hash1} shows the permutation $\pi_1$ which is used to derive $h_1$. For $S_1$ we see that the first $1$ occurs for the index $3$. 
Therefore, we have 
\begin{align*}
& h_1(S_1) = 3, h_1(S_2) = 2, h_1(S_3) = 4.\\ 
& h_2(S_1) = 3, h_2(S_2) = 5, h_2(S_3) = 1,\\
& h_3(S_1) = 1, h_3(S_2) = 5, h_3(S_3) = 1,\\ 
& h_4(S_1) = 1, h_4(S_2) = 5, h_4(S_3) = 1.
\end{align*}

After reorganization, we get four sets given by 
\[S_4 = \{3,2,4\}, S_5 = \{3,5,1\}, S_6 = \{1,5,1\}, S_7 = \{1,5,1\}.\] 

\begin{table}[ht]
\centering
\begin{tabular}{|l|llll|}
\hline
U & \textbf{$S_4$} & \textbf{$S_5$} & \textbf{$S_6$} & \textbf{$S_7$} \\ \hline
1  & 0  & 1     & 1     & 1     \\
2  & 1  & 0     & 0     & 0     \\
3  & 1  & 1     & 0     & 0     \\
4  & 1  & 0     & 0     & 0     \\
5  & 0  & 1     & 1     & 1     \\
6  & 0  & 0     & 0     & 0     \\
7  & 0  & 0     & 0     & 0     \\
8  & 0  & 0     & 0     & 0     \\
9  & 0  & 0     & 0     & 0     \\
10  & 0 & 0     & 0     & 0     \\
\hline
\end{tabular}
\caption{Sets $S_4,S_5,S_6,S_7$ represented using $0-1$ encoding, we take the universal set $U$ and the index set $I$ to be the same since they are  numbers}
\label{table:zoencode2}
\end{table} 
Now \cref{table:zoencode2} shows $0$-$1$ encoding for sets $S_4,S_5,S_6$, and $S_7$. 
We again apply the same hash functions for these sets to obtain the following hashes: 
\begin{align*}
& h_1(S_4) = 3, h_1(S_5) = 3, h_1(S_6) = 1, h_1(S_7) = 1,\\ 
& h_2(S_4) = 4, h_2(S_5) = 5, h_2(S_6) = 5, h_2(S_7) = 5,\\
& h_3(S_4) = 3, h_3(S_5) = 1, h_3(S_6) = 1, h_3(S_7) = 1,\\ 
& h_4(S_4) = 4, h_4(S_5) = 1, h_4(S_6) = 1, h_4(S_7) = 1.
\end{align*}

We finally concatenate them to get the recursive MinHash signature of the three input sets $S_1, S_2$, and $S_3$:
\[ [3,4,3,4,3,5,1,1,1,5,1,1,1,5,1,1].\]

While we use permutations to generate hashes   here, permutations are costly operations to be used in practice. 
Instead, other hashes such as MD5 can be used in practice. 
In addition, the permutations used in this example  were not generated randomly. They are used to  define and illustrate the techniques.  

\section{Quantitative Evaluation}
\label{sec:evaluation}

We provide additional quantitative evaluation results,  including a comparison to a clustering algorithm, a comparison of precision and recall to runtime, and the runtime analysis as a function of LSH parameters $r$ and $k$.

\subsection{Comparison to Clustering Algorithm}

\begin{figure}[!ht]
    \centering
    \includegraphics[width=0.8\linewidth]{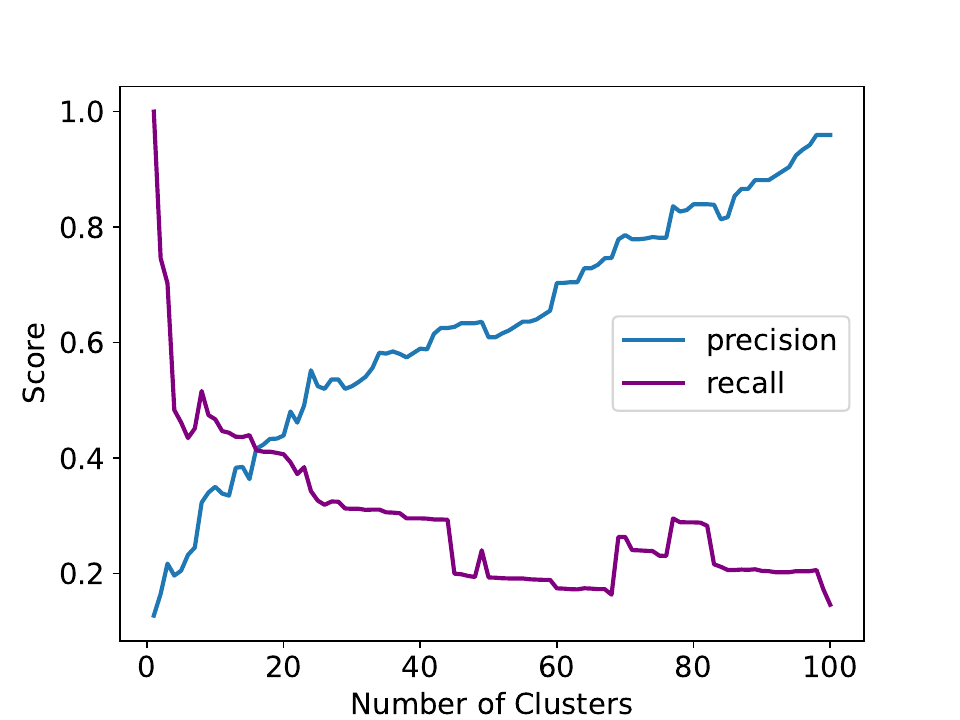}
    \caption{TOSCA shape dataset: precision and recall using K-Medoids clustering. Number of clusters vary from 1 to 100.}
    \label{fig:kmedoids}
\end{figure}

In this section, we provide additional quantitative evaluation on comparison to a clustering algorithm, $k$-medoids. Unlike $k$-means, which requires a well-defined notion of the mean, the $k$-medoids algorithm uses actual data points as the cluster centers. We use the TOSCA shape dataset to compare the precision and recall of each cluster with our results obtained using the LSH methods. We use the merge tree edit distance matrix as input to the $k$-medoids algorithm.~\cref{fig:kmedoids} shows an evaluation of the number of clusters, $n$, and the precision and recall scores using $k$-medoids.

The precision generally increases as $n$ increases because we would get a higher ratio of correctly predicted labels to all the labels in the cluster. Conversely, the recall generally decreases as $n$ increases because the ratio of correctly predicted labels in the same cluster becomes smaller. When $n = 10$, which is the ground truth clustering number in the TOSCA dataset (since we have $10$ different shapes), both precision and recall are below $0.5$. These values are at the same level as both $d_S$ and $d_R$ when $r = 2$, and they are below the scores obtained when $r = 1$.

\subsection{Runtime Analysis}

We perform additional analysis on how runtime directly affects the performance of the LSH methods and the relationship between runtime and LSH parameters, $r$ and $k$.

\begin{figure}[!ht]
    \vspace{-2mm}
    \centering
    \includegraphics[width=0.8\linewidth]{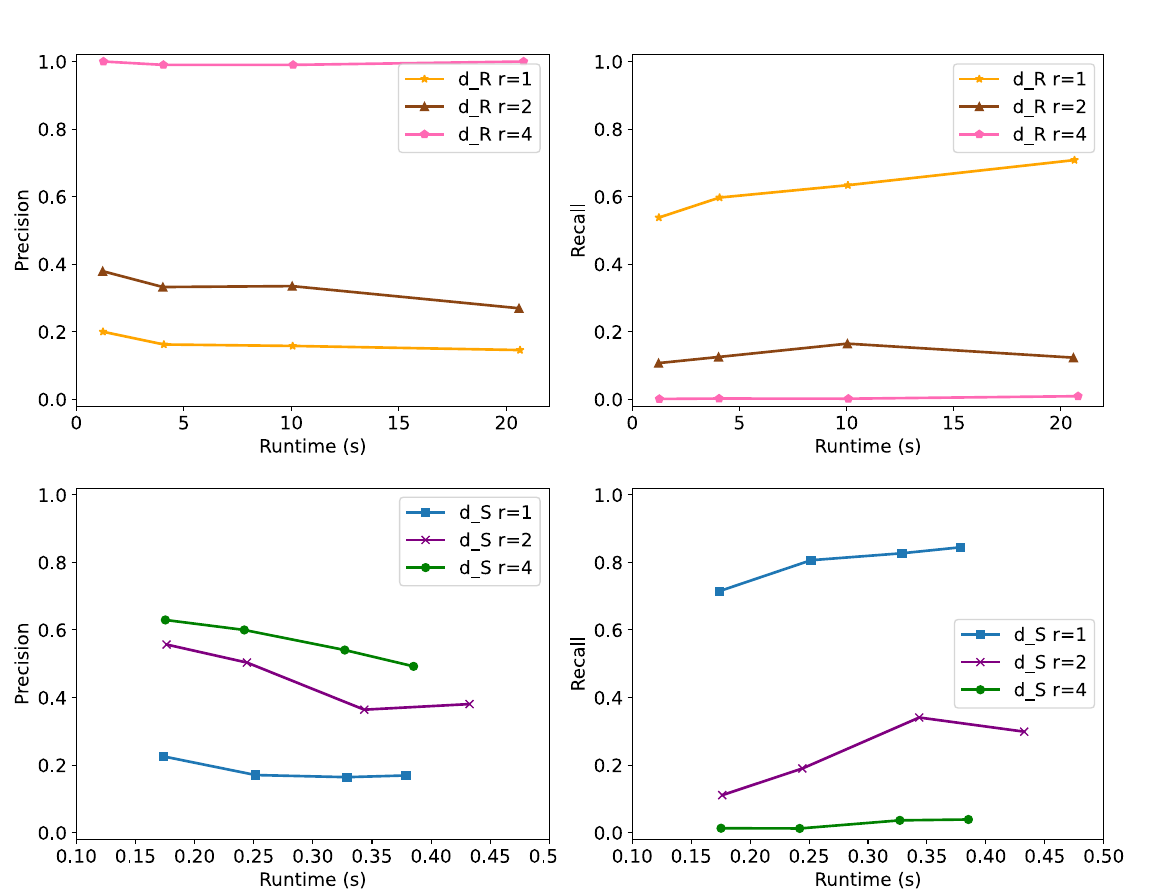}
    \caption{TOSCA shape dataset: runtime vs. precision and recall for both $d_S$ and $d_r$.}
    \label{fig:runtime-vs-pr}
    \vspace{-2mm}
\end{figure}

We investigate the relationship between $k$ and runtime, demonstrating that $k$ can be a representation of runtime, and showing that the trends of runtime vs. precision and recall are similar to those of $k$ vs. precision and recall. In this section, we explicitly provide an evaluation of runtime vs. precision and recall for both $d_S$ and $d_R$ (see~\cref{fig:runtime-vs-pr}). 

The primary computational cost is computing the number of hash functions of each data object, but there are other lower-order terms that do not depend on the size of each object. For instance, the number of rows, $r$, is a lower-order term that does not noticeably affect the runtime. 
We provide \cref{tab:example1} and \cref{tab:example2} to demonstrate that changing $r$ has little affect on the runtime with the same $k$, for both $d_S$ and $d_R$, respectively. 
The runtime remains dominated by $k$. 

\begin{table}[!ht]
    \centering
    \begin{tabular}{|c|c|c|c|c|}
        \hline
        $r$ & $k=20$ & $k=40$ & $k=60$ & $k=80$ \\ \hline
        1 & 0.17375625 & 0.25166304 & 0.32877218 & 0.37872514 \\ \hline
        2 & 0.17633292 & 0.24426496 & 0.34348022 & 0.43249424 \\ \hline
        4 & 0.17529604 & 0.24204648 & 0.32700760 & 0.38517716 \\ \hline
    \end{tabular}
    \caption{$d_S$ for TOSCA shape dataset: r vs. All times are in seconds.}
    \label{tab:example1}
    
\end{table}
\begin{table}[!ht]
    \centering
    \begin{tabular}{|c|c|c|c|c|}
        \hline
        $r$ & $k=20$ & $k=40$ & $k=60$ & $k=80$ \\ \hline
        1 & 1.24099221 & 4.07319212 & 10.06167383 & 20.63537765 \\ \hline
        2 & 1.22652745 & 4.02467799 & 10.03928399 & 20.58727741 \\ \hline
        4 & 1.26209531 & 4.05044847 & 10.08340383 & 20.80203819 \\ \hline
    \end{tabular}
    \caption{$d_R$ for TOSCA shape dataset: $r$ vs. Runtime. All times are in seconds.}
    \label{tab:example2}
\end{table}

\subsection{Additional Visualization Results}

\begin{figure}[!ht]
    \centering
    \includegraphics[width=0.8\linewidth]{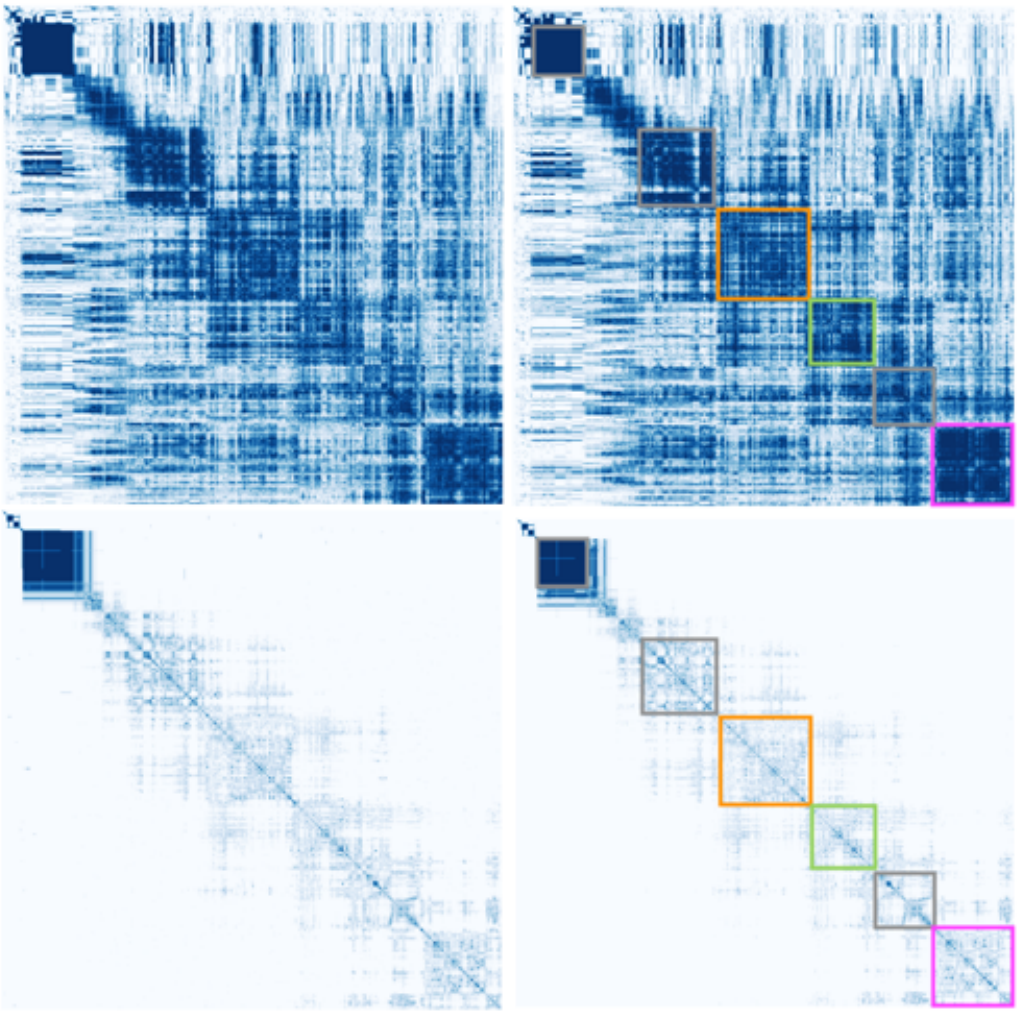}
    \caption{Heated Flow dataset: $d_S$ (top) and $d_R$ (bottom) binary matrices with (right) and without (left) annotated boxes surrounding the clusters.}
    \label{fig:heatedflow}
\end{figure}

We provide additional visualization results of the Heated Flow dataset, showing the original binary matrices without any annotation in~\cref{fig:heatedflow}. 
The clusters are still visible without the annotated boxes.

\end{document}